\documentclass[a4paper,fleqn]{cas-sc}

\ExplSyntaxOn
\cs_gset:Npn \__first_footerline:
  { \group_begin: \small \sffamily \__short_authors: \group_end: }
\ExplSyntaxOff 

\usepackage{amssymb}
\usepackage{enumerate}
\usepackage{mathtools}
\usepackage{graphicx}
\usepackage{caption}
\usepackage{subcaption}
\usepackage{multirow}
\usepackage{adjustbox,lipsum}
\usepackage{tablefootnote}
\usepackage{booktabs}
\usepackage{tabularx}
\usepackage{adjustbox,lipsum}
\usepackage{tablefootnote}
\usepackage{cleveref}
\usepackage{capt-of}
\usepackage{adjustbox}
\usepackage[sort, numbers]{natbib}
\usepackage{lineno}
\usepackage{enumitem}
\usepackage{lipsum}
\usepackage{placeins}

\begin{document}
\let\WriteBookmarks\relax
\def\floatpagepagefraction{1}
\def\textpagefraction{.001}

\shorttitle{Racial and Ethnic Disparities in Transit Supply during the COVID-19 Pandemic}

\shortauthors{Gazmeh et~al.}

\title [mode = title]{Racial and Ethnic Disparities in Transit Supply During the COVID-19 Pandemic: Insights from 232 U.S. Transit Agencies}             



%

\author[1]{Hossein Gazmeh}
\credit{Conceptualization, Methodology, Formal analysis, Writing – Original, Review \& Editing, Visualization}

\author[2]{Lijun Sun}
\credit{Methodology, Software, Data Curation, Supervision}

\author[3]{Yuntao Guo}
\credit{Writing – Review \& Editing, Data Curation, Supervision}

\author[1, 4]{Steven Jones}
\credit{Resources, Writing - Review \& Editing}

\author[1]{Xinwu Qian\corref{cor1}}
\credit{Conceptualization, Formal analysis, Methodology, Writing - Review \& Editing, Supervision}
\cortext[cor1]{E-mail: xinwu.qian@ua.edu}

\affiliation[a]{organization={Department of Civil, Construction and Environmental Engineering, The University of Alabama}, 
            city={Tuscaloosa},
            postcode={35487}, 
            state={AL},
            country={U.S.}}
\affiliation[b]{organization={Department of Civil Engineering, Mcgill University},
            city={Montr\'eal},
            state={Qu\'ebec},
            country={Canada}}
\affiliation[c]{organization={School of Transportation Engineering, Tongji University},
            city={Shanghai},
            country={China}}
\affiliation[d]{organization={Transportation Policy Research Center, The University of Alabama},
            city={Tuscaloosa},
            postcode={35487}, 
            state={AL},
            country={U.S.}}



\begin{abstract}
COVID-19 introduced tremendous disruptions to the normal operation of transit agencies, led to a massive suspension of regular service, and disturbed the access to essential mobility for many transit-dependent populations. In this study, we reexamine the national-level service disruption of the public transportation system as a result of the COVID-19 pandemic and investigate the disparities in the reduction of transit supply, causing certain racial-ethnic groups to be disproportionately affected. To support our analysis, we collect General Transit Feed Specification (GTFS) data from 232 transit agencies covering over 89 million people across 45 states in the U.S. Our findings suggest more disadvantaged communities and certain racial-ethnic groups have experienced disproportional reductions in transit supply, regardless of their pre-pandemic level of service. Further, we employ causal mediation analyses to address the mediation effect of the pre-pandemic level of transit service on the relationship between race/ethnicity and transit supply reduction. This analysis validates the disproportionate reduction in service quality for specific racial/ethnic groups, a trend evident across the majority of transit agencies. In particular, Black Americans are found to be associated with the greatest absolute service loss and also the largest total effects on the transit supply reductions (Black Total Effects = $0.014$), while the White and Asian groups are less impacted (White Total Effects = $-0.018$). We further report that the most impacted communities by transit loss mainly consist of minority populations, where the reductions in transit are correlated with pandemic-related health and economic burdens and result in greater changes in mobility intensity. Our study contributes to a comprehensive understanding of transit inequities due to the pandemic, and provides valuable insights to better prepare public transportation systems against future nationwide disasters.
\end{abstract}




\maketitle

\section{Introduction}
\label{section:intro}
The COVID-19 pandemic led to substantial disruptions in transit services and imposed unprecedented challenges for transit users and agencies to adjust their demands and supply to the new circumstances. On the one hand, there was a consistent decline in ridership due to passengers' concerns about the role of public transportation as a significant disseminator of the virus during its initial takeoff~\citep{qian2021scaling, harris2020subways,qian2021connecting}. At the same time, due to pandemic-induced operation costs and a result of the drastically reduced ridership, extensive service suspensions were observed among transit agencies across the U.S.~\citep{palm2021riders,he2022covid}. For instance, by the end of March 2020, bus and rail ridership in Washington D.C. experienced a sharp decline of 75\% and 90\%, respectively~\cite {wmata}. On the other hand, transit developments in the U.S. are known for their negative contributions to the racial-ethnic disparities in wealth, health, and access to fair opportunities, especially between White and Black Americans~\citep{garrett1999reconsidering, marcantonio2017confronting,larson2018examining}. In particular, public transportation remains an essential mobility mode (if not the only one) for disadvantaged populations, of which the accessibility to daily needs is unevenly distributed during regular services~\citep{stoll2000within}. In the wake of the COVID-19 pandemic, the above issues raise the question of whether the disruptions in transit services have exacerbated disparity gaps in the U.S. and their relation to the challenges faced by the affected communities. 

Prior to the COVID-19 pandemic, a persistent gap existed across racial-ethnic groups in terms of wealth, health and environmental justice. For instance, there remains a large gap of 20 to 30\% between Black and White Americans in homeownership for more than a century~\citep{asante202160}, and the 7-year racial gap in life expectancy between Black and White Americans in 1960 was only reduced to 4 years by 2018~\citep{life_expectancy}. At the same time, the racial and ethnic equity issues are intertwined with the public transportation policies considering the operation and planning of transit services~\citep{pucher1982discrimination}, transit access~\citep{marcantonio2017confronting} and transit-oriented development~\citep{saldana2012racial}. Specifically, studies have shown that transit policies have mainly reinforced gentrification and segregation of communities to more and less transit-accessible areas, impacting the housing prices and quality of access to employment and healthcare~\citep{tehrani2019color, glaeser2008poor, padeiro2019transit}. Moreover, a plethora of evidence indicates that vulnerable populations are often disproportionately hit in times of crisis~\citep{neal2020economic, wheelock2020comparing, reed2012race}, and we hypothesize the loss of transit services during the pandemic is no different. For COVID-19 alone, studies have revealed the gender, racial, and ethnic gaps in job losses~\citep{cowan2020short, montenovo2020determinants}, health risks~\citep{selden2020covid}, and COVID-19 diagnoses and death cases~\citep{Abedi2021, Adams2020, Cajner2020,millett2020assessing}. Furthermore, investigations of passengers' mobility patterns suggested that more economically disadvantaged populations and people of color were inclined to maintain their transit demand despite the pandemic~\citep{sy2021socioeconomic, liu2020impacts, palm2021riders}, and vulnerable populations relied on transit to access essential services and healthcare facilities (e.g., for dialysis)~\citep{nie2022impact}. Nevertheless, the evidence from transit agencies indicates that service adjustments vary in both the magnitude of reduction and their impact on communities with different sociodemographic statuses~\citep{kar2022public, deweese2020tale, wilbur2020impact}.   

In this study, we aim to investigate the possible racial-ethnic disparities in the reduction of transit services during the COVID-19 pandemic. We specifically focus on disparities among racial and ethnic groups, rather than other socioeconomic factors, which is motivated by the fact that the racial and ethnic profile of the transit-dependent population is significantly different than those populations that are less transit-dependent~\citep{Allen2017, neff2007profile}. We quantify disparity as the percentage of transit supply changes as compared to the pre-pandemic level of service, and we adopt a multi-dimensional assessment that accounts for the heterogeneous operational capacity of transit agencies captured by ridership and coverage areas. In addition to the impacts within the transit systems, we further explore the possible health and economic aftereffects on the disproportionately impacted communities that stem from the loss of transit service. In light of all these issues, we propose a series of hypotheses to unveil the potential disparities across different racial-ethnic groups, as summarized below:

\begin{enumerate}[label=\textbf{{H-\bfseries\arabic*}}]
    \item \textit{For certain racial-ethnic groups, the more race/ethnicity-specific population the local community had, the greater the reduction in transit supply would be.}
\end{enumerate}
The first hypothesis gives an understanding of the possible disparities among communities in terms of the absolute quantity of the transit service reduction. However, a high loss during the pandemic may be attributed to a high level of transit supply pre-pandemic. Thereby, we seek to explore the impact of racial-ethnic groups on the reduction of transit supply with regard to the pre-pandemic level. This gives rise to a better understanding of the disparities in the quality of the transit supply, as described by our second hypothesis:

\begin{enumerate}[label=\textbf{{H-\bfseries\arabic*}}, start=2]
    \item \textit{For certain racial-ethnic groups, the more race/ethnicity-specific population the local community had, the greater the reduction in transit supply would be regardless of the pre-pandemic service level.}
\end{enumerate}
Additionally, our third hypothesis aims to reveal if the racial-ethnic disparities in the quality of service reduction can be observed across the majority of transit agencies:

\begin{enumerate}[label=\textbf{{H-\bfseries\arabic*}}, start=3]
    \item \textit{There exists a racial-ethnic disparity across the majority of the transit agencies so that transit-dependent commuters from certain racial-ethnic groups experienced a disproportionate loss in transit access.}
\end{enumerate}
Finally, our fourth hypothesis concerns the potential associations between the pandemic-related health, economic, and mobility intensity burdens with the racial-ethnic disparities in transit service loss:

\begin{enumerate}[label=\textbf{{H-\bfseries\arabic*}}, start=4]
    \item \textit{For the disproportionately impacted racial-ethnic communities, reduction in the transit supply is more correlated with pandemic-related hardships including COVID-19 death rates, job losses, and reductions in mobility intensities.}
\end{enumerate}

To test our hypotheses, we collect data from 232 agencies' operational shifts and the sociodemographic statuses of their 21,714 covered communities at the census tract level. Furthermore, we address \textbf{H-1} by presenting our analysis of the quantity of transit supply change among different racial-ethnic groups. For \textbf{H-2} and \textbf{H-3}, we develop a multi-level causal mediation framework and explore the results both on the national and agency level. This will confirm a racial and ethnic disparity in both the quantity and the quality of the shifts in transit service during the pandemic. Our validation of \textbf{H-4} is based on correlating the shifts in the transit supply with COVID-19 death rates, job losses and changes in mobility behaviors for the most impacted transit-dependent communities on the county level. \Cref{fig:study_framework} presents the study framework.

\begin{center}
\includegraphics[width=.6\linewidth]{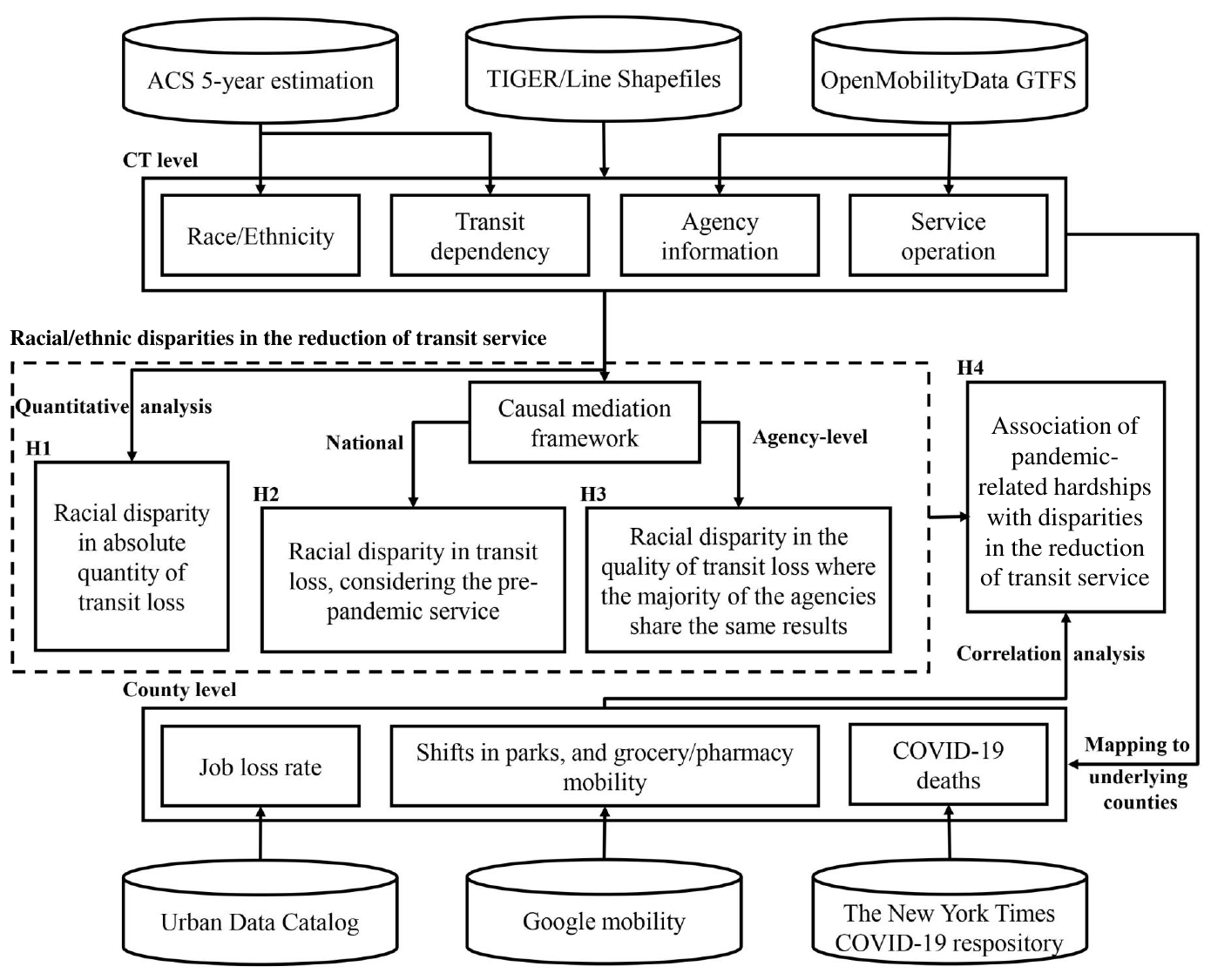}
\captionof{figure}{Study framework}
\label{fig:study_framework}
\end{center}

The rest of this study is organized as follows. First, in \Cref{section:data_collection}, we present the data collection and processing steps used to analyze transit supply reduction resulting from the pandemic, as well as shifts in mobility activities, COVID-19 health and economic implications. \Cref{section:methods} outlines the sociodemographic and agency-related variables employed to assess disparities in the reduction of transit services. We then introduce a causal mediation framework designed to account for pre-pandemic service levels across communities. The results of our primary hypotheses are provided in \Cref{section:results}. Furthermore, in \Cref{section:discussion}, we provide a discussion by summarizing our key findings, conclusions and future direction.

\section{Data Collection and Processing}
\label{section:data_collection}
\subsection{Shifts in Transit Supply}
First, transit agencies' detailed service data, including scheduled transit routes, stops, trips and stop times, were fetched from an open-source service, OpenMobilityData~\citep{TransitFeed2022}, in the form of the General Transit Feed Specification (GTFS)~\citep{gtfs}. Moreover, we narrowed down the selection of agencies to those with valid reports during the pre-pandemic period (from the beginning of 2018 to the conclusion of the first quarter of 2020) and at the initial takeoff of the pandemic (during the second quarter of 2020). Our definition for each period is based on the overall service elimination/restoration trends across the agencies as well as the lockdown orders across the covered states. For each of these periods, we selected the first available GTFS data. As a result, a total of 232 valid transit agencies were identified across 45 states. \Cref{fig:transit_agencies} displays the spatial distribution of the 232 selected agencies. Among the 232 agencies, 122 agencies serve populations of less than 200,000, 87 agencies cover populations ranging from 200,000 to 1,000,000, and the remaining 23 agencies provide transit services to populations exceeding 1,000,000. These classifications align with the categorization of agencies found in the National Transit Database (NTD)~\citep{ntd}, and we designate them as `small', `medium', and `large', respectively. Second, to understand causal factors that are likely contributors to the supply change, we obtain the demographic and economic statuses of the communities across the nation by collecting population, race, income and transportation usage data from the American Community Survey (ACS) 2016-2020 5-year estimates on the census tract (CT) level~\citep{acs5year}. Next, using TIGER/Line geometric datasets, a shapefile containing the above data for all the CTs is created, where all the collected data are processed and consolidated by mapping to the underlying CTs~\citep{tiger}. The covered region by the selected agencies consists of 21,714 CTs with a population exceeding 89 million people. We further report that 9.7\% of the population is served by small agencies, while more than 46\% is covered by medium-sized agencies, and 23 large agencies serve the remaining population of about 39.7 million.

\begin{center}
\includegraphics[width=0.6\linewidth]{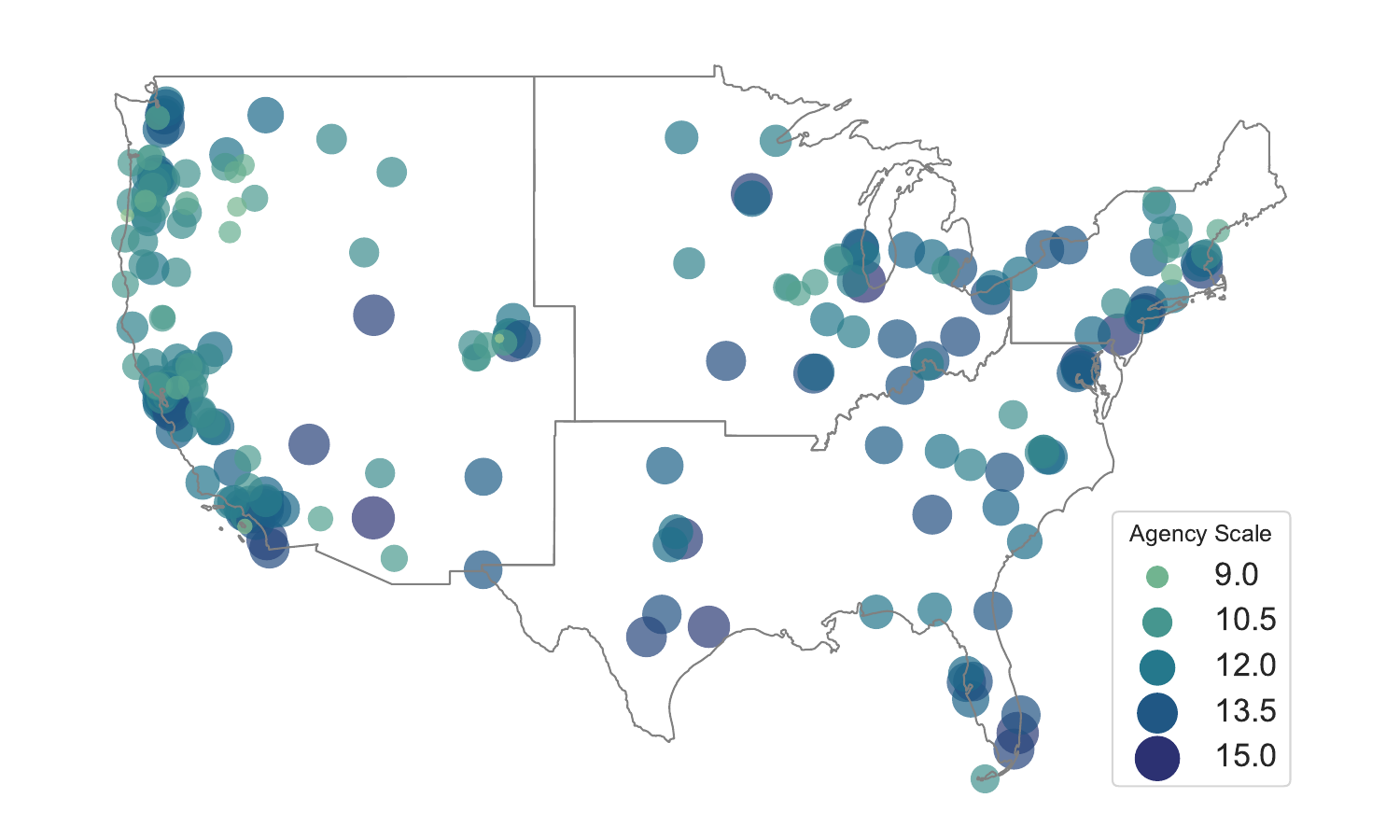}
\captionof{figure}{Spatial distribution of 232 transit agencies}
\label{fig:transit_agencies}
\end{center}

\subsection{COVID-19 Deaths, Job Losses, and Mobility Shifts}
We use The New York Times data repository on cumulative COVID-19 cases and deaths in the U.S., available on the county level~\citep{nyt_covid}. In line with our previous definition of the pandemic period as the second quarter of 2020, we base our analysis on the total number of death cases up to the end of June 2020. The mobility data presented in this study can be accessed via Google COVID-19 Community Mobility Reports~\citep{google_mobility}. The data shows the mobility activities (visits and length of stay at different categories of places) changes compared to a 5-week period (Jan. 3 – Feb. 6) as the baseline. Data includes six categories of places, namely grocery \& pharmacy, parks, transit stations, retail, residential and workplaces. We note that communities vary significantly in their capacities to relocate or remote working options~\citep{brough2021}. Also, shifts in transit stations captured by the big mobility data (such as Google and Apple) can greatly differ from the agencies data, leading to under or over-estimation of the actual trends~\citep{wang2022big}. Thus, we focus on two categories of mobility activities, ``grocery \& pharmacy'' and ``parks'' and use the mean value for mobility changes during the pandemic period. Information on the estimated job loss due to the pandemic comes from Urban Data Catalog~\citep{pandemic_job_loss_data}, capturing the estimated percentage of the job loss till August 2021. Finally, we map all the above information to the 381 counties covered by the 232 selected transit agencies.

\section{Methods}
\label{section:methods}
\subsection{Variables}\label{subsection:measures}
\textbf{Change Rate}: In line with our hypotheses, the dependent variable in our study is the rate of transit service reduction for the covered CTs. For further analysis, the total weekly transit trips (TWTT) are calculated as the total number of trips made at transit stops within a CT throughout a week. This accounts for the scenario where a specific transit route will make multiple stops in a large CT to cater to the travel needs of a greater population. Next, we measure the reduction in CT $i$ transit supply level by calculating the change in TWTT per population (transit supply) between the pre-pandemic and pandemic periods, which we will refer to as the ``Change Rate'', as expressed in~\Cref{eq:change_rate}. A zero value indicates no change of transit supply, and we set the maximum transit supply in the pre-pandemic period at 1, and we cap the maximum change rate at 0.3 to remove excessive outliers such as a transit hub located inside a commercial area with few populations. This leads to a total of 21,714 observations, where the mean change rate is 0.031 with a standard deviation of 0.054. 

\begin{equation}
    \begin{aligned}
        \text{Change Rate}_{i} = \frac{\text{TWTT}_{i,\text{pre-pandemic}} - \text{TWTT}_{i,\text{pandemic}}}{\text{Population}_i}
    \end{aligned}
    \label{eq:change_rate}
\end{equation}

\textbf{Race/Ethnicity}: Moreover, we are primarily interested in racial and ethnic status for independent variables. We focus on the three major races in the U.S.: White, Black, and Asian. We also account for people of Hispanic or Latino ethnicity. More specifically, consistent with Census Bureau terms, we refer to ``non-Hispanic White'' as ``White'', ``non-Hispanic Black or African American'' as ``Black'', ``non-Hispanic Asian'' as ``Asian'' and ``people with Hispanic or Latino origin'' as ``Hispanic''. Since the calculation of the change rate already eliminates the effect of the population size of the CTs, we also quantify race/ethnicity impacts by measuring their composition over the entire population. In our study, the investigated transit service areas consist of 48.6\% of White, 15.4\% of Black, 7.6\% Asian and 24.4\% of Hispanic. This reveals a lower representation of White individuals (as opposed to 58.9\%) and a higher representation of Black (compared to 13.6\%), Asian (as opposed to 6.3\%), and Hispanic individuals (as opposed to 19.1\%) compared to the national average. 

\textbf{Transit Dependency and Agency Scale}: In addition to race/ethnicity, it is also important to acknowledge the impacts of potential confounder variables that could be linked to both race/ethnicity and changes in transit supply. For example, economic status and car ownership are two likely confounders that impact both the variables~\citep{chong2006experiences,gautier2010car}. Since there are a plethora of factors that may build a bridge between race/ethnicity and the transit supply, here we simply use the baseline transit dependency as the single indicator of all such factors. This transit dependency is captured in the ACS as the percentage of the population in a CT who use public transportation for work commuting. In our study areas, transit dependency has an average of 0.071 and a standard deviation of 0.118. We also consider an agency-level confounder denoted by the ``Agency Scale'' (mean=13.54, std=1.02). The variable is calculated as the logarithmic value of the total covered population of each transit agency, which serves as a proxy of the agency's size and coverage area. \Cref{tab:stat} presents the summary statistics of the variables.

\begin{table}[H] 
\centering 
\small
\caption{Summary Statistics (N=21,714)}   
\label{tab:stat} 
\begin{tabular}{@{\extracolsep{5pt}}llllllll}
\textbf{Statistic} & \multicolumn{1}{c}{\textbf{Mean}} & \multicolumn{1}{c}{\textbf{St. Dev.}} & \multicolumn{1}{c}{\textbf{Min}} & \multicolumn{1}{c}{\textbf{25\%}} & \multicolumn{1}{c}{\textbf{50\%}} & \multicolumn{1}{c}{\textbf{75\%}} & \multicolumn{1}{c}{\textbf{Max}} \\
\midrule
\textbf{Dependent}\\
Change Rate  & 0.031 & 0.054 & 0.000 & 0.000 & 0.004 & 0.040 & 0.300 \\ [1.2ex]
\textbf{Race/Ethnicity}\\
White & 0.486 & 0.292 & 0.000  & 0.216 & 0.515 & 0.746 & 1.000 \\ 
Black & 0.154 & 0.230 & 0.000  & 0.015 & 0.054 & 0.172 & 1.000 \\ 
Asian & 0.076 & 0.113 & 0.000  & 0.009 & 0.034 & 0.093 & 0.904 \\ 
Hispanic & 0.244 & 0.248 & 0.000  & 0.057 & 0.146 & 0.357 & 1.000 \\  [1.2ex]
\textbf{Transit related}\\
Pre-pandemic Supply & 0.152 & 0.179 & 0.000  & 0.025 & 0.087 & 0.211 & 0.996 \\ 
Agency Scale & 13.544 & 1.023 & 7.629  & 13.003 & 13.622 & 14.350 & 15.049 \\ 
Transit Dependency & 0.071 & 0.118 & 0.000  & 0.006 & 0.026 & 0.079 & 1.000 \\
\bottomrule \\[-1.8ex] 
\end{tabular} 
\end{table}

\subsection{Causal Mediation Analysis}\label{section:cma}
Most of the efforts in recent decades to examine the linear causal relationships between the observed and latent variables have been based on path analytical models, particularly Structural Equation Modeling (SEM)~\citep{baron1986moderator, mackinnon2007mediation}. However, a number of shortcomings in the traditional SEM framework have been raised recently. For instance, the conventional SEM framework is not generalizable to nonlinear problems and also falls short of offering a comprehensive definition of causal mediation effects as its identification assumptions are model-specific. At the same time, advancements in causal mediation analysis have emerged as a response to the shortcomings of traditional SEM, where causal mediation analysis can be regarded as a broader framework that encompasses SEM as a special case~\citep{imai2010general, vanderweele2015explanation, posch2021testing}. As a result, it has lately seen extensive applications in different domains of social science, psychology, and epidemiology~\citep{vanderweele2016mediation, mackinnon2007mediation, richiardi2013mediation}. In our study, causal mediation analysis enables us to establish a common framework for the definition, identification and estimation of causal relationships between the treatment (X) and the outcome (Y) through the mediator (M). Here, X corresponds to the composition of racial-ethnic groups, Y is the change rate, and M concerns the transit supply level before the pandemic in a CT. In line with our discussions, \Cref{fig:cma_framework} shows the proposed framework.  

\begin{center}
\includegraphics[width=.6\linewidth]{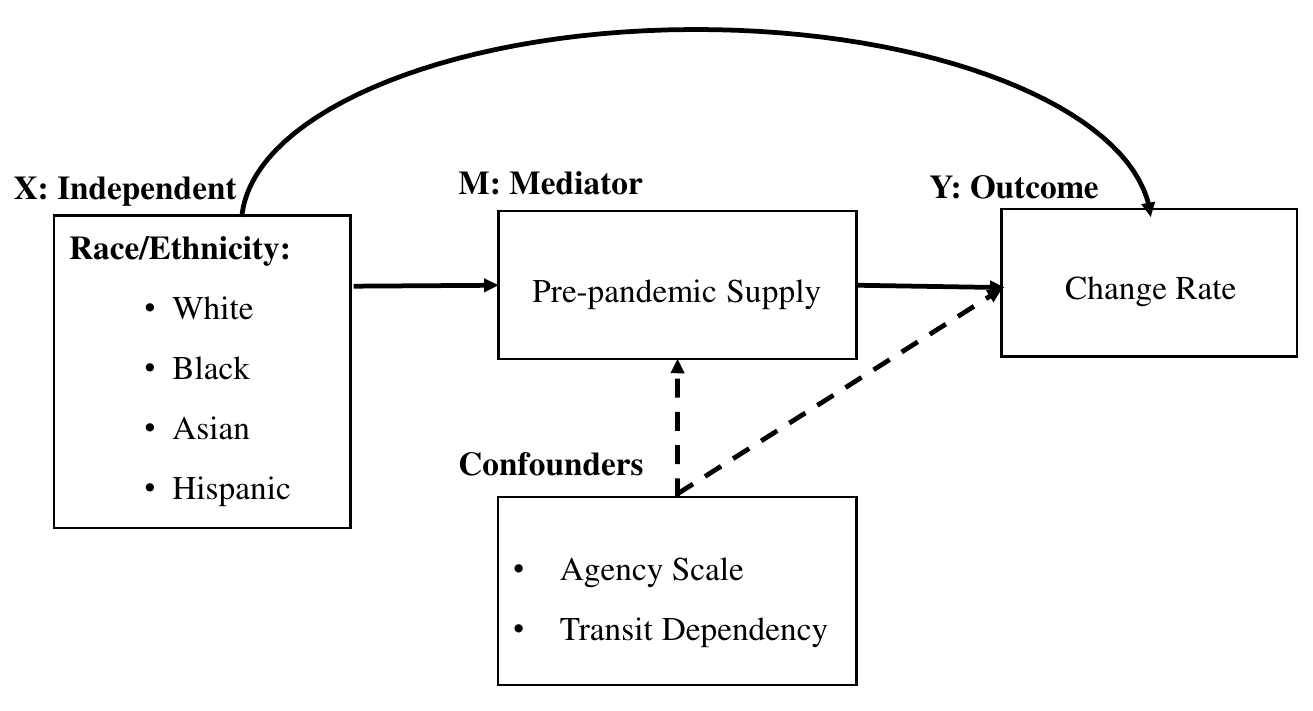}
\captionof{figure}{Conceptual causal mediation framework}
\label{fig:cma_framework}
\end{center}

Although conventional path analytical models, such as SEM, can separate the direct effect \((X\xrightarrow{}Y)\) from the indirect effect \((X\xrightarrow{}M\xrightarrow{}Y)\) and lead to an interpretation of possible racial and ethnic disparities in the reduction of transit service, we suspect that their modeling assumption on the path independence between \(X\xrightarrow{}M\) and \(M\xrightarrow{}Y\) is likely violated. \Cref{fig:scatter_nonlinear} displays the relationship between the race/ethnicity percentages (X) and pre-pandemic level of transit supply (M) at different values of change rate. Each axis is divided into ten quantiles, and the mean change rate for each range is indicated by the color density. We observe that the impact of the pre-pandemic supply (M) on change rate (Y) is not independent of the racial-ethnic percentage (X). In particular, for the Black group, a high level of transit service before the pandemic will more likely lead to a high change rate in CTs with a relatively low or high composition of Black Americans. Thereby, to achieve an unbiased understanding of both direct and indirect effects, the causal mediation analysis is employed to address the interdependency between the causal pathways and the nonlinear interactions among the variables.

\begin{center}
\includegraphics[width=1\textwidth]{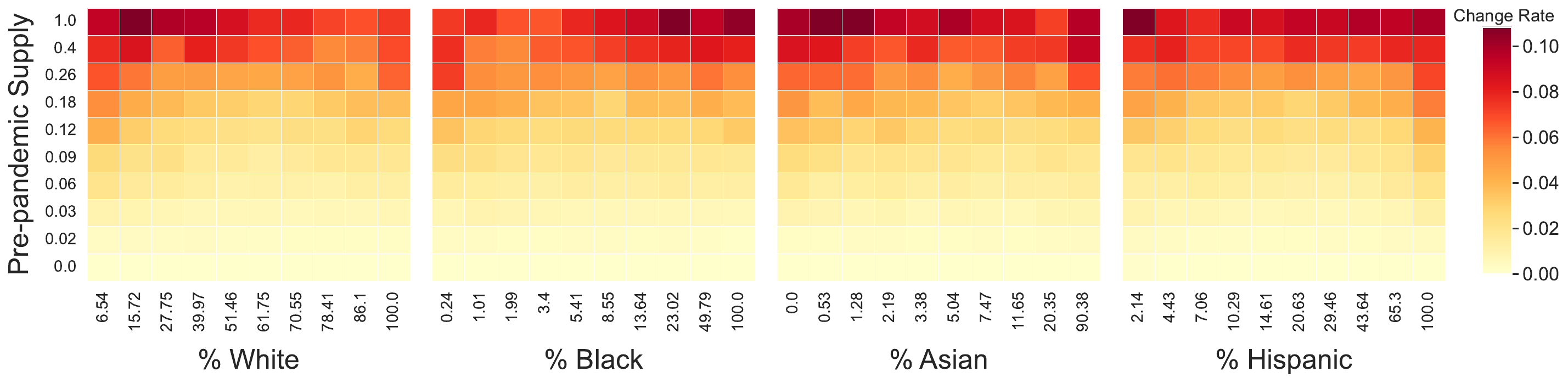}
\captionof{figure}{The relationship between the CT's racial-ethnic composition and pre-pandemic transit supply for different values of change rate.}
\label{fig:scatter_nonlinear}
\end{center}

The essence of the causal mediation analysis resides in the counterfactual framework to identify the direct and indirect effects, ultimately yielding the total effect. In the context of our study, consider \(Y_i(x, M(x))\) as a function of the change rate in CT \(i\) given the racial-ethnic composition \(x\) and pre-pandemic supply \(M(x)\). The impacts of \(X\) on \(Y\) can be decomposed as the average causal mediation effect (ACME) and the average direct effect (ADE), where the latter captures the direct effects of the racial-ethnic composition on the change rate regardless of the pre-pandemic service. In view of the counterfactual framework, the ACME concerns the indirect effect as the expectation of the function \(\delta_i(x)\) on how Y will change by altering M (e.g., change from high pre-pandemic supply to low pre-pandemic supply) while holding X constant, as presented in \Cref{eq:acme}. Note that for the particular example, \(Y_{i}(x, \text{High Pre-pandemic Supply})\) is observed while \(Y_{i}(x,\text{Low Pre-pandemic Supply})\) is the counterfactual outcome. Although the latter cannot be observed directly, the value of which can always be predicted by estimating the \(Y\sim M + X + X\times M\) from the data (e.g., a regression model).

\begin{equation}
    \small
    \begin{aligned}
    \delta_{i}(x) \equiv Y_{i}(x, \text{High Pre-pandemic Supply})-Y_{i}(x, \text{Low Pre-pandemic Supply})
    \end{aligned}
    \label{eq:acme}
\end{equation}

Similarly, taking a CT with a high percentage of the race/ethnicity-specific population as an example, the ADE concerns the expectation of the function \(\zeta_{i}(x)\) on how \(Y\) will change by altering the race/ethnicity percentage while holding the pre-pandemic supply constant. This is displayed in \Cref{eq:ade}. Once more, the former expression on the right-hand side of the equation is observed, while the latter term is the counterfactual outcome. The counterfactual outcome can also be predicted by the same estimation equation as before.

\begin{equation}
\small
    \begin{aligned}
    \zeta_{i}(x) \equiv Y_{i}(\text{High Race Percentage}, M_{i}(x))-Y_{i}(\text{Low Race Percentage}, M_{i}(x))
    \end{aligned}
    \label{eq:ade}
\end{equation}

Finally, based on ACME and ADE, one can compute the total effect (TE) as \(\mathbb{E}[\delta_i(x)+ \mathbb{E}[\zeta_i(x)]\). Furthermore, considering the fact that the change rate is a result of the shifts in services provided by transit agencies with different capabilities, a multilevel structure is needed to address the heterogeneity across agencies. To this end, we use hierarchical linear modeling (HLM) to fit the mediator and outcome models. For each racial-ethnic group, a variety of linear and polynomial models are examined, and the choice of model forms is based on their goodness of fit. The models for White and Black groups are expressed as follows:

\begin{equation}
\small
\label{eq:mediator}
\begin{aligned}
    \text{Pre-pandemic Supply}_{ij}= \gamma_{0j}+\gamma_{1j}* \text{Race}_{ij}+ \gamma_2*\text{Race}_{ij}^2+&\gamma_3*\text{Agency Scale}_j
    + \gamma_4*\text{Transit Dependency}_{ij} + \epsilon_{ij} &
    \\     \gamma_{0j}=\alpha_{0}+\mu_{0j}, \quad
     \gamma_{1j}=\alpha_{0}+\mu_{1j}
\end{aligned}
\end{equation}

\begin{equation}
\small
\label{eq:outcome}
\begin{aligned}
    \text{Change Rate}_{ij}= \kappa_{0j}+\kappa_{1j}*\text{Pre-pandemic Supply}_{ij}+&\kappa_{2}* \text{Race}_{ij}+  \kappa_{3}* \text{Race}_{ij}^2 + \kappa_{4}*\text{Pre-pandemic Supply}_{ij}*\text{Race}_{ij} \\
    + \kappa_{5}\text{Transit Dependency}_{ij}+ & \kappa_{6} \text{Agency Scale}_j+ \epsilon_{ij} \\    \kappa_{0j}=\alpha_{0}+\mu_{0j},\quad \kappa_{1j}=\alpha_0+\mu_{1j}
\end{aligned}
\end{equation}

For the Asian group, the HLM model for the mediator is the same as Equation \eqref{eq:mediator} while the model for the Hispanic group excludes the term \(\gamma_2*\text{Race}_{ij}^2\). On the other hand, the outcome HLM model for both Asian and Hispanic groups is identical where there is no quadratic term \(\kappa_{3}* \text{Race}_{ij}^2\) as Equation \eqref{eq:outcome}. In all models, predictors are scaled and centered around the mean. We also allow random intercepts for each transit agency \(j\) to account for the unobserved heterogeneity. In other words, adding the random intercept considers the agencies' differences in providing/eliminating the service. The random slopes are introduced for race/ethnicity in the mediator model and for the pre-pandemic level of supply in the outcome model. The former addresses the demographical differences between the covered areas by the agencies, while the latter means that the rate at which the change rate is impacted by the pre-pandemic level of service can be different across the agencies. Finally, the variables agency scale and transit dependency are used as the two confounders at the agency level and the observation level, and the nonlinear interaction between the racial-ethnic composition and the pre-pandemic supply is introduced in the outcome model. Given the finalized HLM models, the model parameters are estimated using the lmer package in R~\citep{lmer_r} for the HLM models and the mediation package in R~\citep{tingley2014mediation} for the causal mediation effects.

\section{Results}
\label{section:results}
\subsection{Racial-ethnic Disparity in the Quantity of Transit Supply Reduction}\label{section:h1}
Our initial findings delve into the socioeconomic and racial-ethnic attributes of the covered CTs, categorized by the absolute magnitude of transit supply reduction. This gives us a better understanding of the disparities regarding the quantity of the reduction in transit supply, where the pre-pandemic level of service is not explicitly included. We categorize CTs into four groups based on their loss of transit supply, i.e., change rate: the top 25\% with the highest change rates are labeled as ``High," the next 25\% as ``Med-High," the subsequent 25\% as ``Med-Low," and the lowest 25\% which have a change rate value of 0 are labeled as ``No Change." \Cref{fig:change_socio} shows an apparent disparity between CTs with different socioeconomic statuses in terms of their change rate. More precisely, on average, 12.2\% of the households in CTs with a high change rate do not own a vehicle. This is 15\% higher than the average across all the covered CTs. Moreover, there is a notable gap between the CTs in terms of health insurance coverage. On average, CTs with a high change rate are 45\% more likely to lack coverage compared to those with no change (values being 11.5\% and 7.9\%, respectively). In addition, a higher median income is consistently attributed to a lower change rate (the average median income for CTs with no change is almost 18\% more than CTs with a high change rate). Also, the relative values for transit-dependent populations are similar to people without a vehicle, validating our usage of transit dependency as a representation of various socioeconomic characteristics. More interestingly, the racial and ethnic status of the impacted CTs displays a distinct opposite trend between the Black and White communities regarding the level of change rate (\Cref{fig:change_race}). The average proportion of White people in CTs with no reduction is above 52.2\%, which is 11\% more than their average in CTs with a high change rate. At the same time, CTs with a high change rate consist of 22.2\% Black Americans on average, which is 44.2\% higher than their actual representation in the covered CTs. On the other hand, it is evident that there are no noticeable variations in the CTs' Asian and Hispanic compositions across different change rate levels. \Cref{fig:dependency_race} shows the transit supply levels among racial-ethnic groups prior to the pandemic. From \Cref{fig:dependency_race}, we report that densely populated Black CTs are heavily reliant on the transit system, contrary to the White group. More importantly, the findings from \Cref{fig:change_race} and \Cref{fig:dependency_race} highlight the potential influence of the pre-pandemic supply on the disparities in change rates, given the significant variation across different racial-ethnic groups' pre-pandemic transit supply level. Nevertheless, the qualitative results of the impacted CTs' sociodemographic status confirm our first hypothesis, \textbf{H-1}, regarding the disproportionate transit loss in quantity for certain transit-dependent racial-ethnic groups. However, these findings also prompt the question of whether the observed disparities remain when accounting for the pre-pandemic level of transit supply.  
\begin{center}
\includegraphics[width=0.85\textwidth]{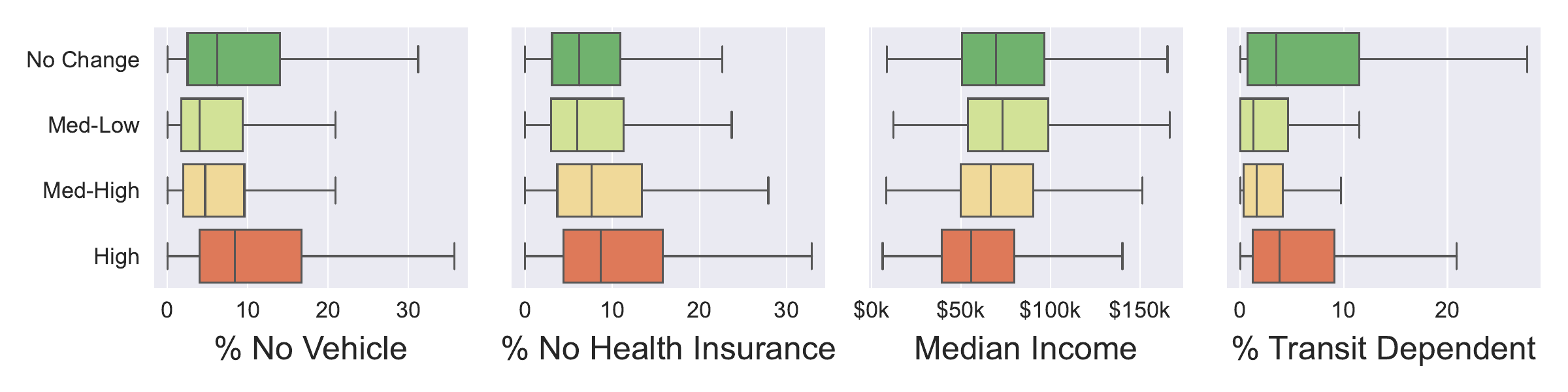}
\captionof{figure}{Socioeconomic status across different levels of transit change}
\label{fig:change_socio}
\end{center}
\begin{center}
\includegraphics[width=0.85\textwidth]{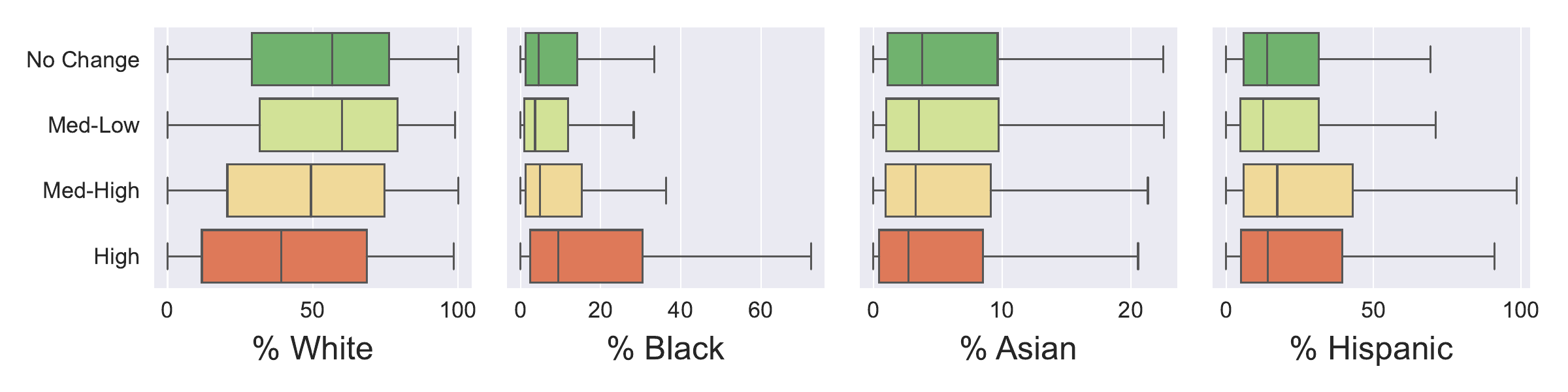}
\captionof{figure}{Racial-ethnic status across different levels of transit change}
\label{fig:change_race}
\end{center}
\begin{center}
\includegraphics[width=0.85\textwidth]{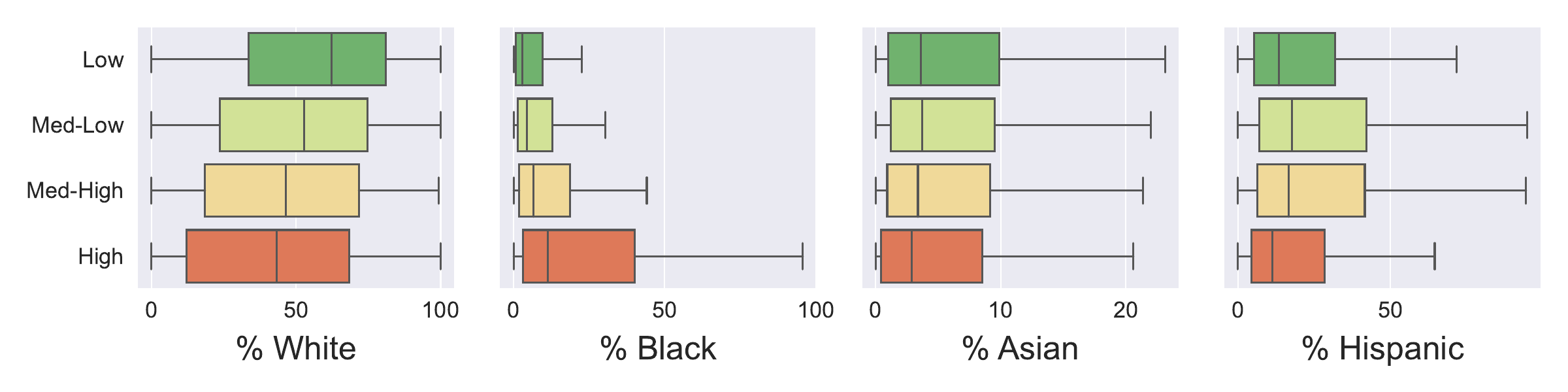}
\captionof{figure}{Racial-ethnic status across different levels of pre-pandemic transit supply}
\label{fig:dependency_race}
\end{center}

\subsection{Racial-ethnic Disparity in the Quality of Transit Supply Reduction}
To account for the role of pre-pandemic transit supply on the observed transit loss disparities among racial-ethnic groups, we employ a causal mediation framework, outlined in section \eqref{section:cma}. We present the results of the model in the following order. First, an initial hierarchical linear model examines the significance of independent variables (racial-ethnic compositions) and transit-related confounders (agencies' scale, transit dependency) in explaining the mediator (pre-pandemic supply levels). Specifically, our focus lies in determining whether the racial-ethnic composition of CTs \((X)\) has a significant impact on the pre-pandemic transit supply level \((M)\). Subsequently, another hierarchical linear model connects the race/ethnicity \((X)\), pre-pandemic supply \((M)\) and transit-related confounders to the change rate \((Y)\). The main question here revolves around whether the interaction of racial-ethnic compositions and pre-pandemic supply significantly influences the change rate, confirming the mediation effect of the pre-pandemic service level. Finally, we break down the impact of race/ethnicity on the change rate into average \textbf{direct} effects and average causal mediation, \textbf{indirect}, effects and provide a discussion of the results for each racial-ethnic group.

\begin{table}[]
\centering
\small
\caption{Hierarchical Linear Model Results for Pre-pandemic Supply.}
\label{table:x_to_m}
\begin{tabular}{@{\extracolsep{5pt}}l l l l l}
     & \textbf{White} & \textbf{Black} & \textbf{Asian} & \textbf{Hispanic}\\
    \midrule
    (Intercept)             & $0.078^{*}$        & $0.084^{*}$ & $0.089^{*}$ & $0.091^{*}$ \\
                            & $(0.035)$      & $(0.034)$     & $(0.036)$     & $(0.035)$     \\
    Transit Dependency      & $0.567^{***}$  & $0.537^{***}$ & $0.575^{***}$ & $0.570^{***}$ \\
                            & $(0.016)$      & $(0.015)$     & $(0.016)$     & $(0.015)$     \\
    Agency Scale            & $-0.009^{**}$  & $-0.010^{**}$ & $-0.010^{**}$ & $-0.010^{**}$ \\
                            & $(0.003)$      & $(0.003)$     & $(0.003)$     & $(0.003)$     \\
    \% White             & $-0.083^{***}$ &               &               &               \\
                            & $(0.013)$      &               &               &               \\  
    \% White$^2$             & $0.033^{.}$ &               &               &               \\
                            & $(0.019)$      &               &               &               \\  
    \% Black             &                & $0.094^{***}$ &               &               \\
                            &                & $(0.010)$     &               &               \\
    \% Black$^2$             &                & $0.050^{*}$ &               &               \\
                            &                & $(0.021)$     &               &               \\
    \% Asian          &                &               & $-0.098^{***}$       &               \\
                            &                &               & $(0.016)$     &               \\
    \% Asian$^2$          &                &               & $0.136^{*}$       &               \\
                            &                &               & $(0.057)$     &               \\
    \% Hispanic             &                &               &               & $-0.034^{***}$      \\
                            &                &               &               & $(0.007)$     \\
    \midrule
    AIC                     & $-21490.507$   & $-21615.753$  & $-21482.016$  & $-21466.241$  \\
    BIC                     & $-21418.641$   & $-21543.887$  & $-21410.150$  & $-21402.360$  \\
    ICC                         &$0.100$              &$0.092$        & $0.105$          & $0.102$ \\
    Pseudo-$R^2$ (total)      &$0.280$              &$0.286$        & $0.316$          & $0.311$ \\
    No. Observations               & $21714$        & $21714$       & $21714$       & $21714$       \\
    No. Groups     & $232$          & $232$         & $232$         & $232$         \\
    \bottomrule
    \multicolumn{5}{l}{\scriptsize{$^{***}p<0.001; ^{**}p<0.01; ^{*}p<0.05; \Dot{}p<0.1$}}
\end{tabular}
\end{table}

\Cref{table:x_to_m} presents the results of the first hierarchical linear model, where we explore the influence of the independent variable (racial-ethnic composition) and confounders (transit dependency and agency scale) on the pre-pandemic level of transit supply. For all racial and ethnic groups, the transit dependency of the covered CTs positively and significantly affects the pre-pandemic service level ($p < 0.001$). This indicates that a higher proportion of transit-dependent population in the CTs corresponds to a greater level of transit service prior to the pandemic. The agency scale is found negative and statistically significant ($p< 0.01$), suggesting that the size of the agencies has a negative effect on the trip density of their covered CTs. Additionally, all the linear and quadratic effects of racial-ethnic compositions in our models are found to be statistically significant. Overall, interpreting the linear terms indicates that the proportion of White Americans in a given CT significantly influences the supply level adversely ($p< 0.001$). The same holds for Asian Americans ($p< 0.001$) and linearly for Hispanics ($p< 0.001$). On the other hand, the ratio of the Black population is found to be positive and statistically significant ($p< 0.001$), such that a higher percentage of Black Americans is attributed to more transit supply levels. Additionally, all the quadratic terms are significant and positive, suggesting a convex relationship between the White, Black and Asian proportion in CTs with the pre-pandemic level of supply. In short, the results for the first model emphasize the significant and nonlinear relationships between the percentage of racial-ethnic groups and transit-related variables with the pre-pandemic transit supply, underlining the necessity of capturing the disparities from a causal mediation standpoint. 

\begin{table}[htbp]
\centering
\small
\caption{Hierarchical Linear Model Results for Change Rate.}
\label{table:m_to_y}
\begin{tabular}{@{\extracolsep{5pt}}l l l l l}
 & \textbf{White} & \textbf{Black} & \textbf{Asian} & \textbf{Hispanic}\\
\midrule
(Intercept)             & $-0.047^{**}$        & $-0.046^{**}$ & $-0.047^{**}$ & $-0.047^{**}$ \\
                        & $(0.015)$      & $(0.015)$     & $(0.015)$     & $(0.015)$     \\
Transit Dependency      & $0.020^{***}$  & $0.019^{***}$ & $0.018^{***}$ & $0.018^{***}$ \\
                        & $(0.003)$      & $(0.003)$     & $(0.003)$     & $(0.003)$     \\
Agency Scale            & $0.006^{***}$  & $0.006^{***}$ & $0.006^{***}$ & $0.006^{***}$ \\
                        & $(0.001)$      & $(0.001)$     & $(0.001)$     & $(0.001)$     \\
Pre-pandemic Supply            & $0.225^{***}$  & $0.225^{***}$ & $0.224^{***}$ & $0.224^{***}$ \\
                        & $(0.015)$      & $(0.015)$     & $(0.015)$     & $(0.015)$     \\
\% White             & $0.002^{**}$ &               &               &               \\
                        & $(0.001)$      &               &               &               \\  
\% White$^2$             & $0.001^{.}$ &               &               &               \\
                        & $(0.003)$      &               &               &               \\
\% White $\times$ Pre-pandemic Supply                           & $0.018^{***}$    &                &               &  \\
         & $(0.005)$  &                &               &   \\  
\% Black             &                & $-0.004^{**}$ &               &               \\
                        &                & $(0.001)$     &               &               \\
\% Black$^2$             &                & $0.015^{***}$ &               &               \\
                        &                & $(0.003)$     &               &               \\
\% Black $\times$ Pre-pandemic Supply          &               & $-0.023^{***}$       &   &               \\
                        &          & $(0.005)$     &               &\\
\% Asian          &                &               & $0.002$       &               \\
                        &                &               & $(0.002)$     &               \\
\% Asian $\times$ Pre-pandemic Supply          &                &               & $0.071^{***}$       &               \\
                        &                &               & $(0.013)$     &               \\
\% Hispanic             &                &               &               & $-0.002^{*}$      \\
                        &                &               &               & $(0.001)$     \\
\% Hispanic $\times$ Pre-pandemic Supply            &                &               &               & $-0.019^{**}$      \\
                        &                &               &               & $(0.007)$     \\                        
\midrule
AIC                     & $-97079.017$   & $-97088.785$  & $-97098.620$  & $-97081.451$  \\
BIC                     & $-96991.181$   & $-97000.949$  & $-97018.769$  & $-97001.600$  \\
ICC                         &$0.555$              &$0.555$        & $0.554$          & $0.555$ \\
Pseudo-$R^2$ (total)      &$0.830$              &$0.830$        & $0.830$          & $0.830$ \\
No. Observations               & $21714$        & $21714$       & $21714$       & $21714$       \\
No. Groups     & $232$          & $232$         & $232$         & $232$         \\
\bottomrule
\multicolumn{5}{l}{\scriptsize{$^{***}p<0.001; ^{**}p<0.01; ^{*}p<0.05; \Dot{}p<0.1$}}
\end{tabular}
\end{table}

We present the results of the hierarchical linear model for the change rate in \Cref{table:m_to_y}. We report the effects of transit dependency on the change rate to be positive and statistically significant ($p< 0.001$), indicating a higher reduction of service for CTs with a higher level of transit dependency. Across all racial-ethnic groups, the impact of the pre-pandemic level of service is significantly positive ($p< 0.001$), suggesting a higher pre-pandemic supply led to a greater reduction in the transit service. In addition, larger transit agencies seem to have a more challenging time maintaining their level of service as their operational scale is positively associated with more reduction ($p< 0.001$), which is likely due to the higher level of service before the pandemic. Moreover, the interaction term between racial-ethnic percentages and pre-pandemic supply is significant across all the models (for White, Black and Asian $p< 0.001$, for Hispanic $p<0.01$), confirming the impact of race-ethnicity on change rate to be mediated via the pre-pandemic supply. More interestingly, we observe a notable disparity in the influence of race/ethnicity on the change rate. Specifically, the influence of the proportion of Black or Hispanic Americans is negative and significant (for Black $p< 0.01$, for Hispanic $p<0.05$). At the same time, the composition of White Americans is significantly associated with more reduction in the transit service ($p< 0.01$) and the impact of Asian Americans is found to be positive but insignificant. Nevertheless, capturing an accurate estimation of the impact of race-ethnicity on the change rate requires a causal mediation analysis that accounts for both the direct and indirect effects, which are mediated by the pre-pandemic supply. 

Based on the results from the mediator (pre-pandemic supply) and outcome (change rate) models, we proceed with the causal mediation analysis to unveil the indirect and direct effects of racial-ethnic groups on transit supply reduction. \Cref{fig:mediation} displays the graphical summary of the causal mediation analysis results for racial-ethnic groups, including the average causal mediation effects (ACME) and average direct effects (ADE). Similar to our findings on the absolute quantity of service change in section \eqref{section:h1}, we observe distinct effects for various racial-ethnic groups. Specifically, the indirect effect of transit loss for Black Americans, mediated by the pre-pandemic level of service, is significant and positive (ACME $= 0.019, p< 0.001$), highlighting a connection between transit supply and change rate. On the contrary, the indirect effect of White Americans on the change rate is negative and statistically significant (ACME $= -0.020, p< 0.001$). On the other hand, the direct effect of Black Americans is statistically significant and negatively associated with change rate (ADE $= -0.005, p< 0.001$), whereas a positive and significant effect for White Americans is revealed (ADE $= 0.002, p< 0.05$). However, a notably higher indirect effect resulted in a significant positive total effect for the Black group (Total Effects $= 0.014, p< 0.001$) and a total negative effect for White Americans (Total Effects $= -0.018, p< 0.001$). Asian and Hispanic are found to be the groups with both negative direct and indirect effects. Asian group's direct effect is insignificant (ADE $= -0.001$) while their average causal mediation effect is the highest among all the groups, with a negative and significant value (ACME $= -0.025, p< 0.001$). Consequently, their total effect is found to be negative and statistically significant (Total Effects $= -0.026, p< 0.001$). For Hispanics, the direct effect is negative and significant (ADE $= -0.002, p<0.05$), as the indirect effect is also significant (ACME $= -0.007, p<0.001$), resulting in a negative and significant total effect (Total Effects $= -0.009, p<0.001$). 

\begin{center}
\includegraphics[width=0.6\textwidth]{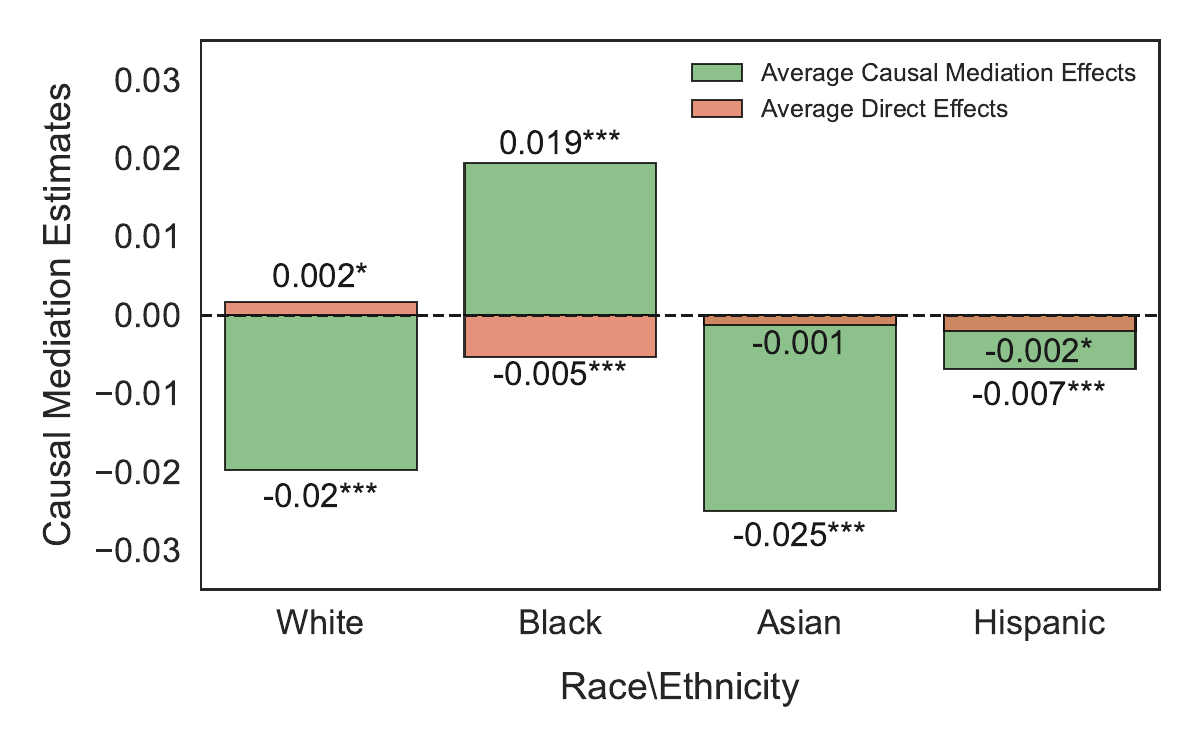}
\captionof{figure}{Results for causal mediation effects.}
\label{fig:mediation}
\end{center}

The above discussions reveal a significant disparity between racial-ethnic groups in the quality of transit service reduction during the COVID-19 pandemic, where the differences in the pre-pandemic levels of transit supply are taken into account. In brief, the indirect effects of race/ethnicity on the change rate highlight the pronounced reliance of Black Americans on the pre-pandemic transit service, as their effect on the change rate is positively and significantly mediated through the pre-pandemic supply. Conversely, the opposite pattern is observed for White, Asian, and Hispanic Americans. Furthermore, the direct effect of the racial-ethnic composition of CTs on the change rate, regardless of their pre-pandemic service levels, indicates a negative direct effect for Black and Hispanic Americans but a positive direct effect for White Americans (with the results for Asian group being insignificant). This can be interpreted as service preservation is generally observed in areas with higher proportions of Black and Hispanic minorities, suggesting possible prioritization of communities with vulnerable populations when adjusting the service. Lastly, the total effects of these racial-ethnic groups on the change rate demonstrate that Black Americans experience the most substantial impact, while White and Asian Americans are comparatively less affected. Hence, based on the results of causal mediation analysis, our second hypothesis (\textbf{H-2}) is confirmed, as considering the pre-pandemic service level highlights a significant disparity in the reduction of transit supply across racial-ethnic groups.

\subsection{Racial-ethnic Disparity in Transit Supply Reduction at Agency Levels}
\begin{figure}[htbp]
    \centering
    \subfloat[White]{\includegraphics[width=0.22\linewidth]{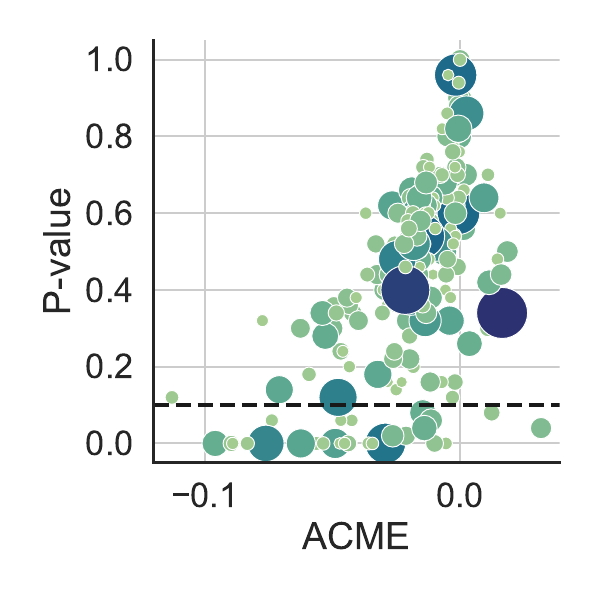}}
    \subfloat[Black]{\includegraphics[width=0.22\linewidth]{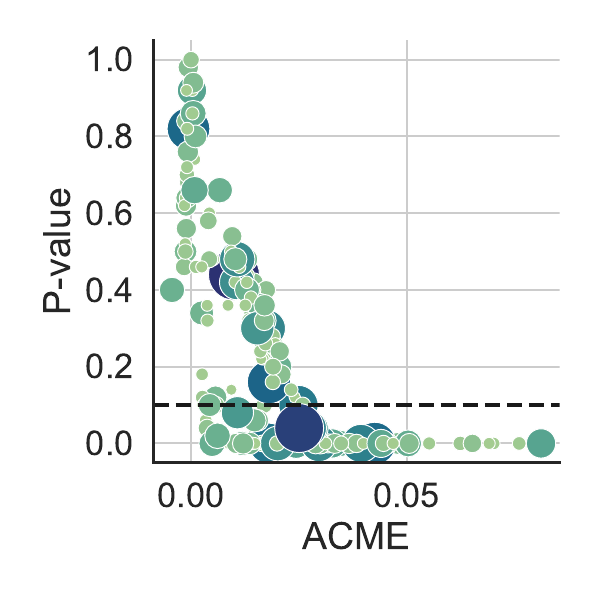}}
    \subfloat[Asian]{\includegraphics[width=0.22\linewidth]{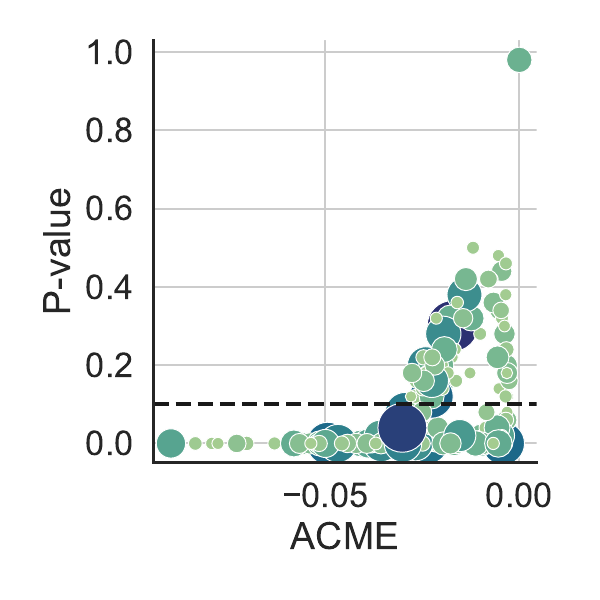}}
    \subfloat[Hispanic]{\includegraphics[width=0.22\linewidth]{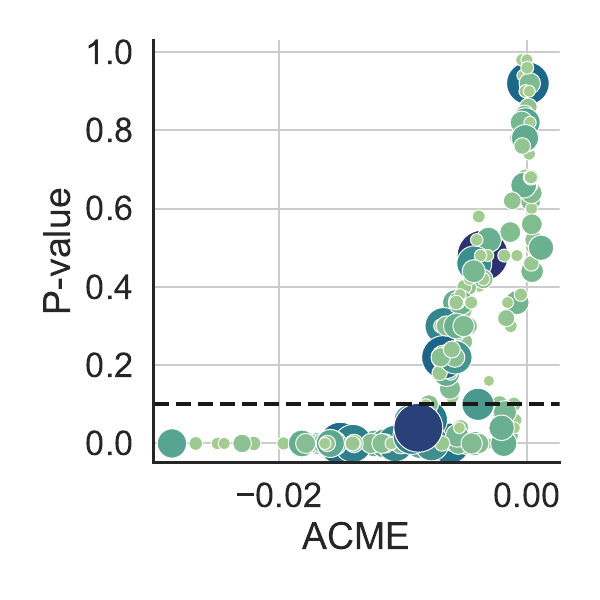}}\\
    \subfloat[White]{\includegraphics[width=0.22\linewidth]{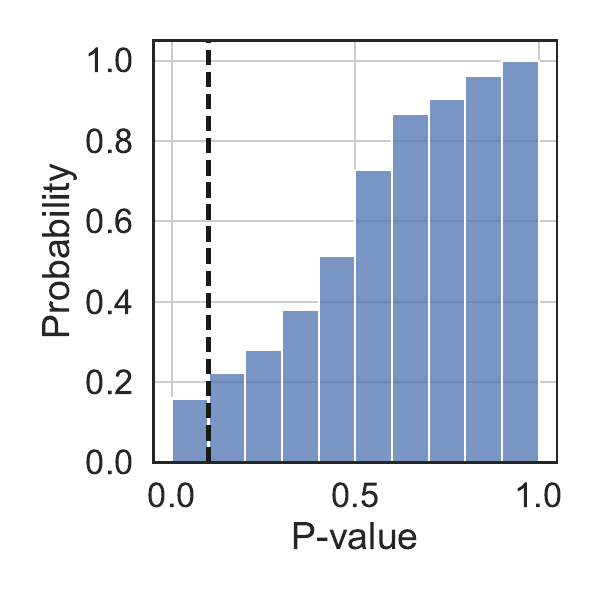}}
    \subfloat[Black]{\includegraphics[width=0.22\linewidth]{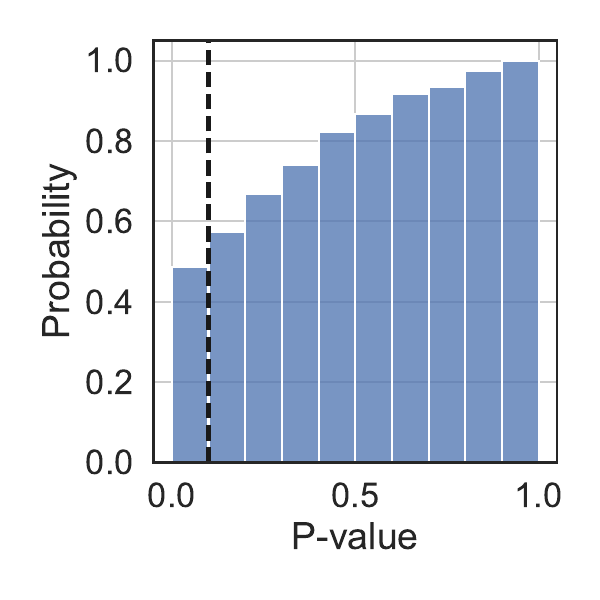}}
    \subfloat[Asian]{\includegraphics[width=0.22\linewidth]{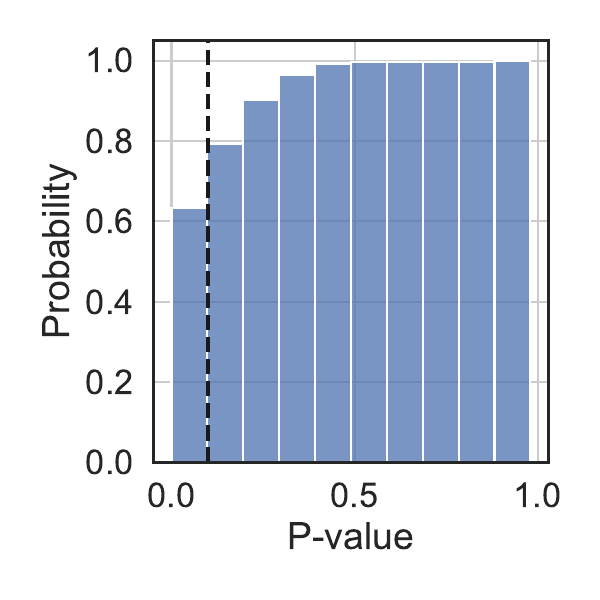}}
    \subfloat[Hispanic]{\includegraphics[width=0.22\linewidth]{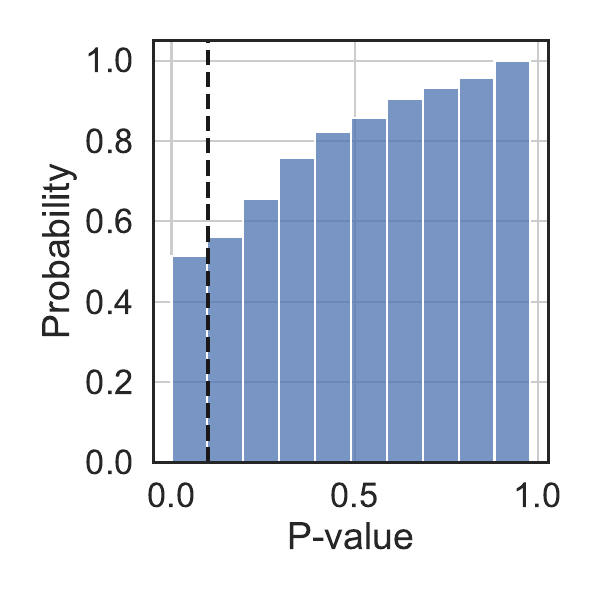}}
    \caption{Estimation results for agency-level average causal mediation effects.}
    \label{fig:agency_acme}
\end{figure}

\begin{figure}[htbp]
    \centering
    \subfloat[White]{\includegraphics[width=0.22\linewidth]{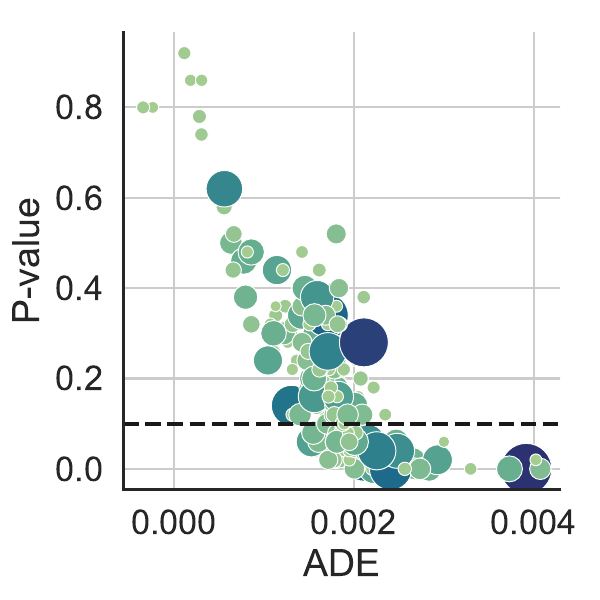}}
    \subfloat[Black]{\includegraphics[width=0.22\linewidth]{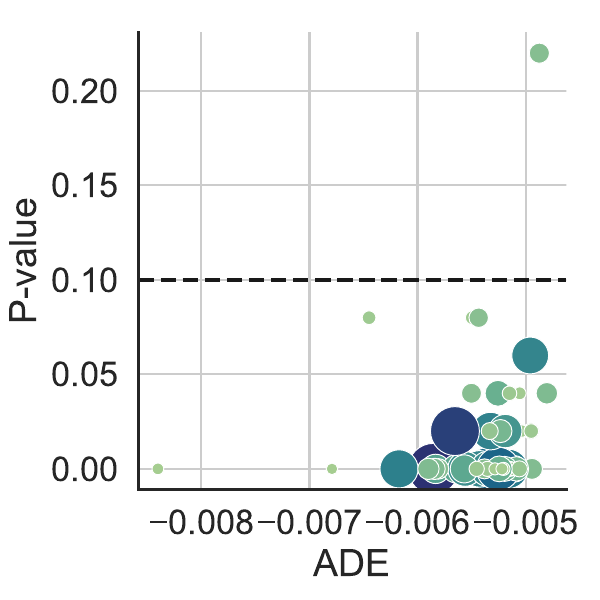}}
    \subfloat[Asian]{\includegraphics[width=0.22\linewidth]{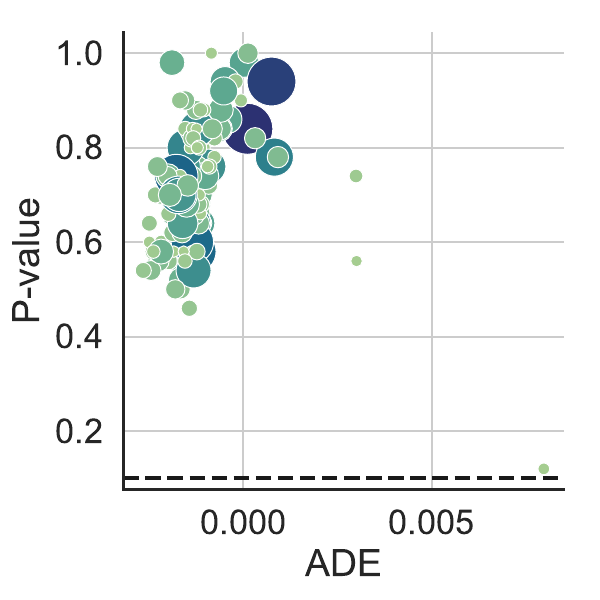}}
    \subfloat[Hispanic]{\includegraphics[width=0.22\linewidth]{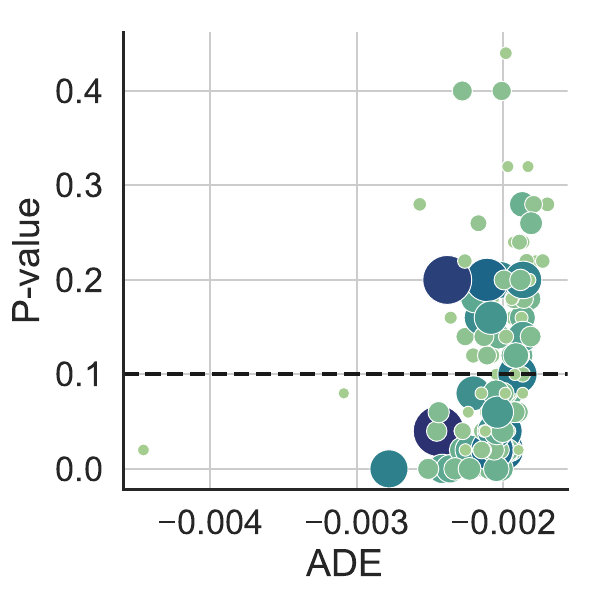}}\\
    \subfloat[White]{\includegraphics[width=0.22\linewidth]{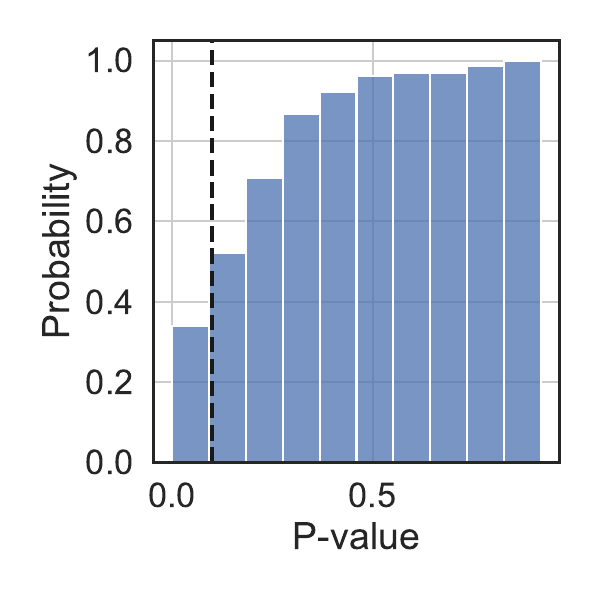}}
    \subfloat[Black]{\includegraphics[width=0.22\linewidth]{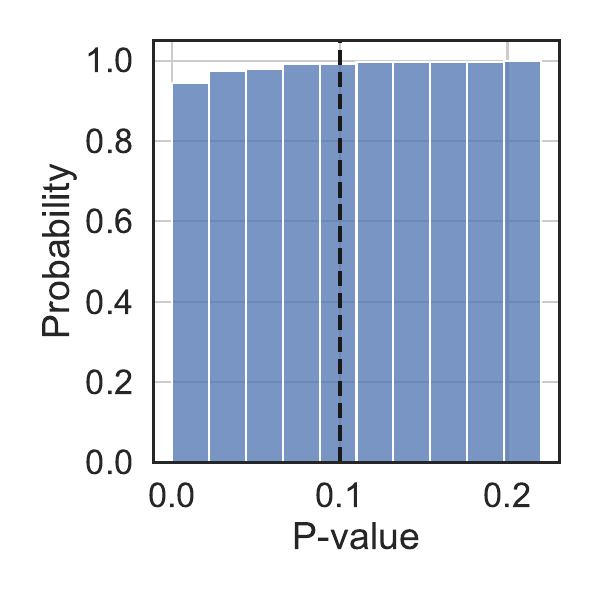}}
    \subfloat[Asian]{\includegraphics[width=0.22\linewidth]{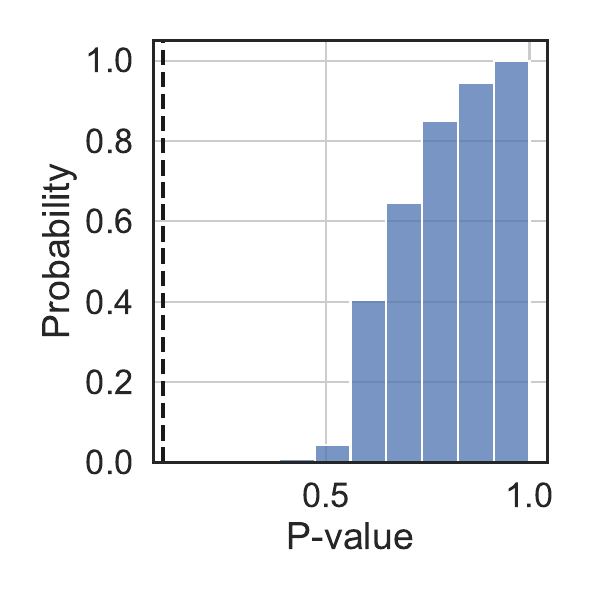}}
    \subfloat[Hispanic]{\includegraphics[width=0.22\linewidth]{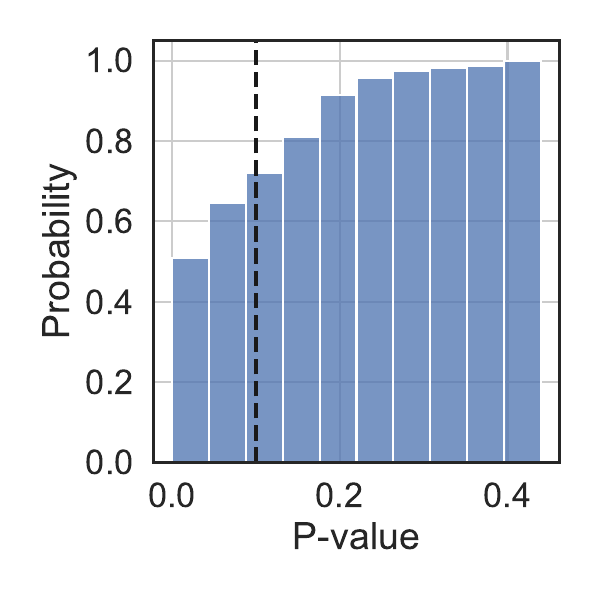}}
    \caption{Estimation results for agency-level average direct effects.}
    \label{fig:agency_ade}
\end{figure}

We are further interested in checking whether the observed disparities represent a common trend among the majority of transit agencies in their service adjustments during the pandemic. We present the results for agency-level causal mediation effects in \Cref{fig:agency_acme} and \Cref{fig:agency_ade}. A threshold (\(p=0.1\)) for the significance of the effects is specified. A slightly higher significance level allows a more inclusive consideration of potential effects and associations, aligning with the aim of capturing a balance between minimizing false negatives and guarding against unwarranted conclusions. Each transit agency is represented by a scatter point (darker and larger points for agencies with a higher value for agency scale).

A review of agency-level direct and causal mediation effects demonstrates two notable trends for shifts in Black Americans' transit supply. First, the pre-pandemic level of service by the agencies is significantly ($p< 0.001$) and positively influenced by the proportion of Black Americans (\Cref{table:x_to_m}), also noticeable from \Cref{fig:agency_acme} (b) with their significant ($p< 0.1$) impact on over 40\% of agencies' ACME. Second, The direct effect of the Black group on supply reduction across all agencies is negative, with more than 95\% being significant (\Cref{fig:agency_ade} (b)). In contrast, while the significant ($p< 0.001$) negative influence of the White group on the pre-pandemic service level already highlighted their less reliance on the transit system  (\Cref{table:x_to_m}), \Cref{fig:agency_ade} (a) displays the positive direct effect of the White composition on the service reduction for all the agencies, with more than 50\% being statistically significant. For the Asian group, the indirect effect is negative across all agencies, while their direct effect is non-significant (\Cref{fig:agency_ade} (c)). Furthermore, the earlier discussion on (\Cref{table:m_to_y}) also highlighted an insignificant contribution of this group to the change rate. These observations collectively indicate that the Asian group appears to be the least reliant and least affected among minority groups. In addition, the results at the agency level indicate a resemblance between the direct effects of the Hispanic group on the change rate and those of the Black group (\Cref{fig:agency_ade} (d), with a negative value across all the agencies (70\% being significant). However, when considering their mediated effect through the pre-pandemic supply level on the change rate, the Hispanic group's pattern aligns more closely with that of the White and Asian groups, with predominantly negative values (\Cref{fig:agency_acme} (d)). The discussed findings confirm a similar pattern among transit agencies regarding the effects of race-ethnicity on observed service reductions, with a largely consistent positive or negative impact on the transit service reductions across the transit agencies. 

We conduct further investigation to explore whether transit agencies of varying sizes differ significantly in terms of the effects of racial-ethnic groups on their change rate. This will highlight whether having access to sufficient resources may mitigate the disparity of transit losses among different racial-ethnic groups. To this end, \Cref{fig:cme_agency_size} presents the causal effects of race-ethnicity compositions on the change rate categorized by agency sizes. The classification of agency sizes follows the convention in the NTD, where agencies are categorized as small if they cover a population of less than 200,000, medium if their coverage ranges from 200,000 to 1,000,000, and large if they serve more than 1,000,000 people~\citep{ntd}. We use the Kruskal-Wallis (KW)~\citep{kruskal1952use} test to examine the null hypothesis that the mean rank of the race-ethnicity causal effects is the same for small, medium, and large transit agencies. This is to account for the non-normality of both direct and indirect causal effects distributions (Shapiro-Wilk~\citep{shapiro1965analysis} test of normality; null hypothesis rejected with $p<0.001$ for all categories), which prevents us from using an ANOVA test. Following the KW test, we fail to reject the null hypothesis for all twelve scenarios covering different combinations of three different causal effects and four racial-ethnic groups (with $p>0.1$). This indicates that transit agency sizes, which represent their operational capabilities and service coverage, may not impact their immediate response to national emergencies such as the pandemic, and the transit riders in small, medium, and large urban areas suffered a similar level of transit supply reduction disparities.

\begin{center}
\includegraphics[width=1\textwidth]{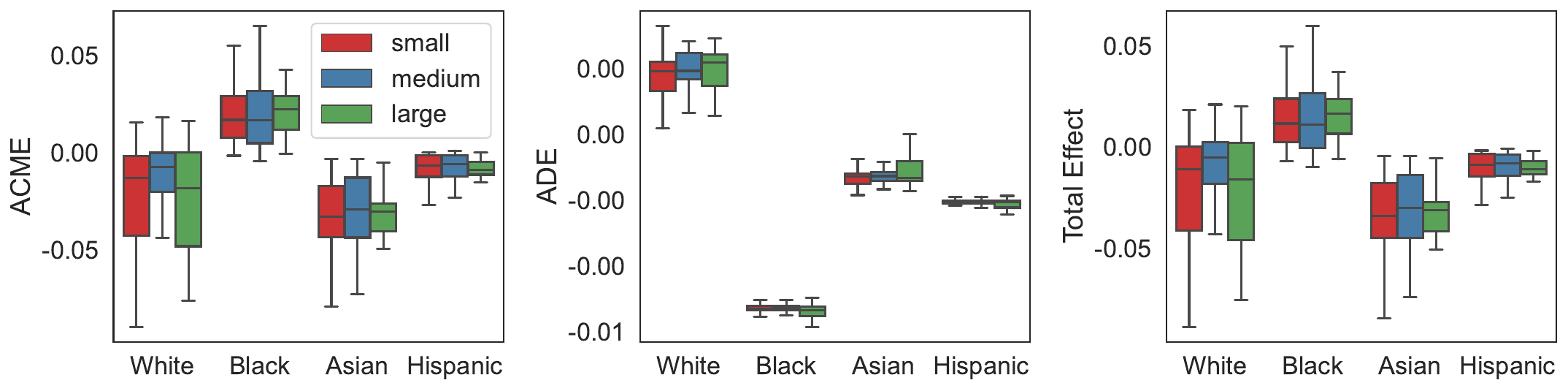}
\captionof{figure}{Distribution of causal effects by agency size for each racial-ethnic group}
\label{fig:cme_agency_size}
\end{center}

Regarding our above discussions on the agency-level causal mediation effects of different racial-ethnic groups on change rates, we observe the disparities to be mainly similar across the majority of the transit agencies. Note that in all the cases related to the direct effects of race/ethnicity, we report a comparable positive or negative influence of race-ethnicity on transit loss across at least 50\% of the transit agencies. In short, our findings verify our third hypothesis, \textbf{H-3}, as there exist similar patterns among transit agencies toward the reduction of their service such that certain racial-ethnic groups are disproportionately impacted.

\subsection{Transit Supply Reduction and the Related Consequences}
\begin{figure}[htbp]
    \centering
    \subfloat[Black]{\includegraphics[width=0.25\linewidth]{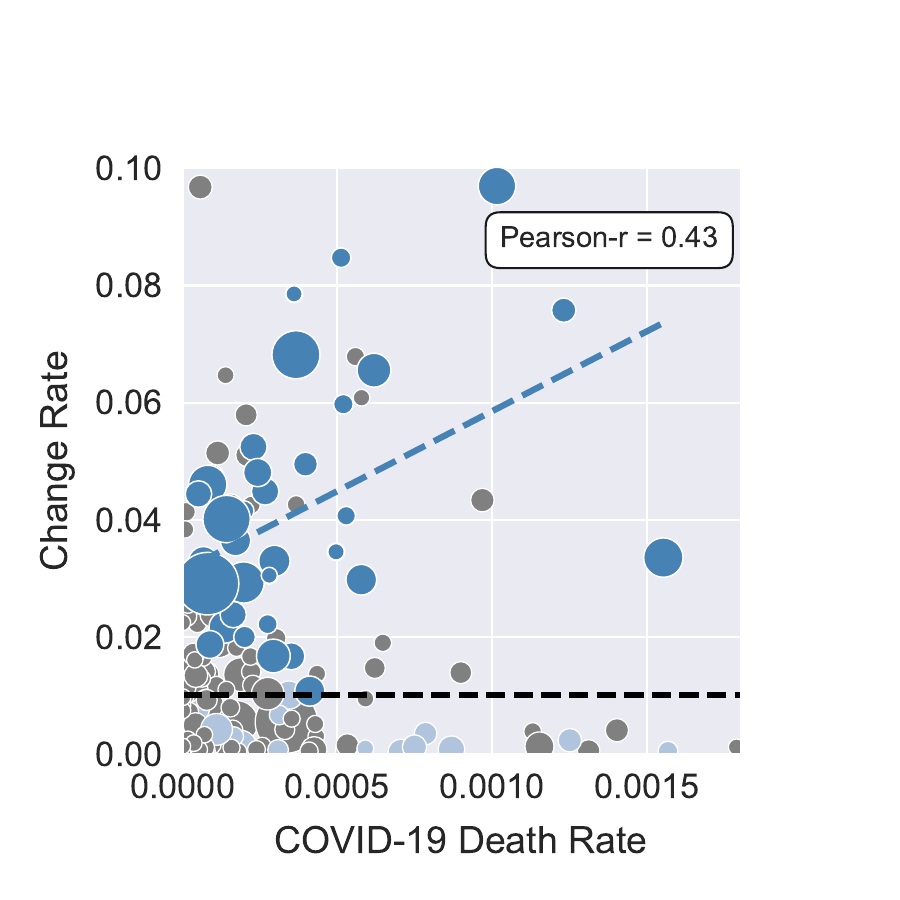}}
    \subfloat[Asian]{\includegraphics[width=0.25\linewidth]{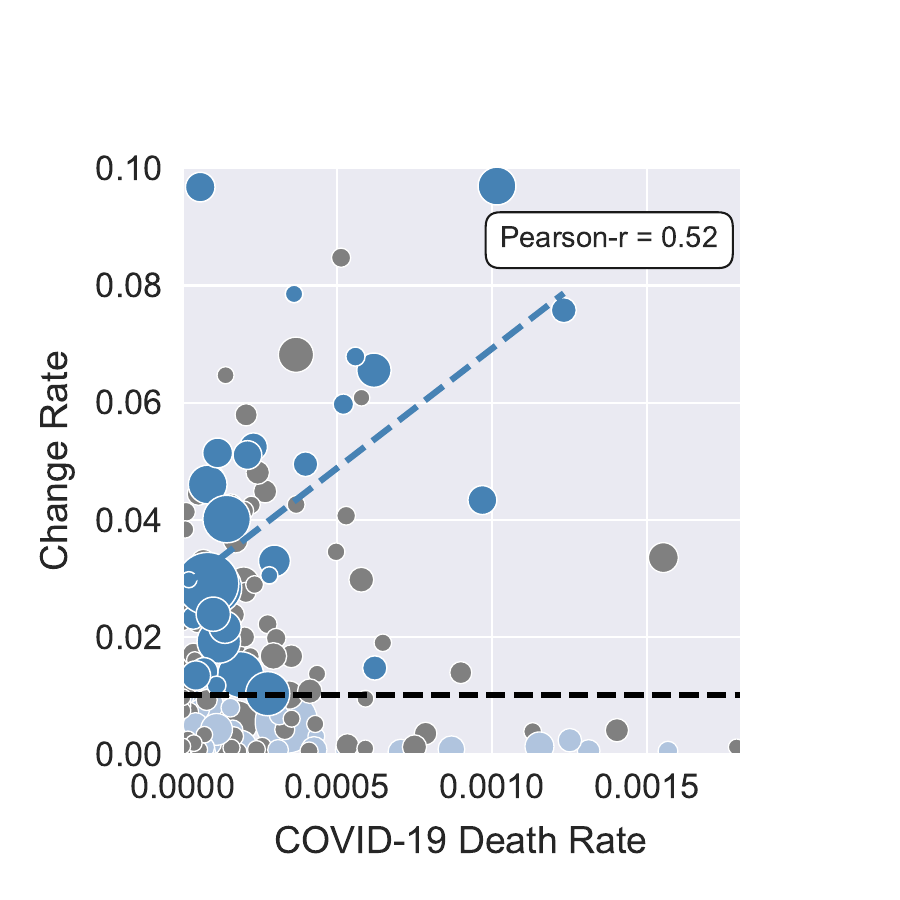}}
    \subfloat[Hispanic]{\includegraphics[width=0.25\linewidth]{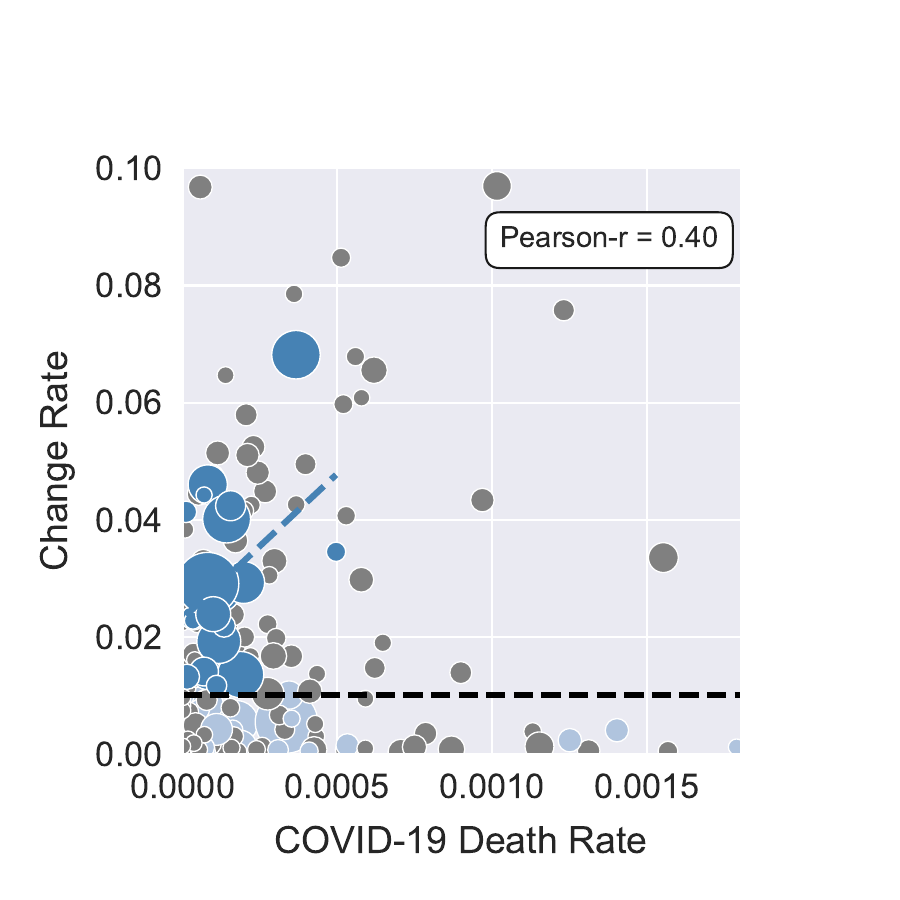}}
    \caption{County-level results for COVID-19 death rates by the end of June 2020.}
    \label{fig:county_covid}
\end{figure}

\begin{figure}[htbp]
    \centering
    \subfloat[Black]{\includegraphics[width=0.25\linewidth]{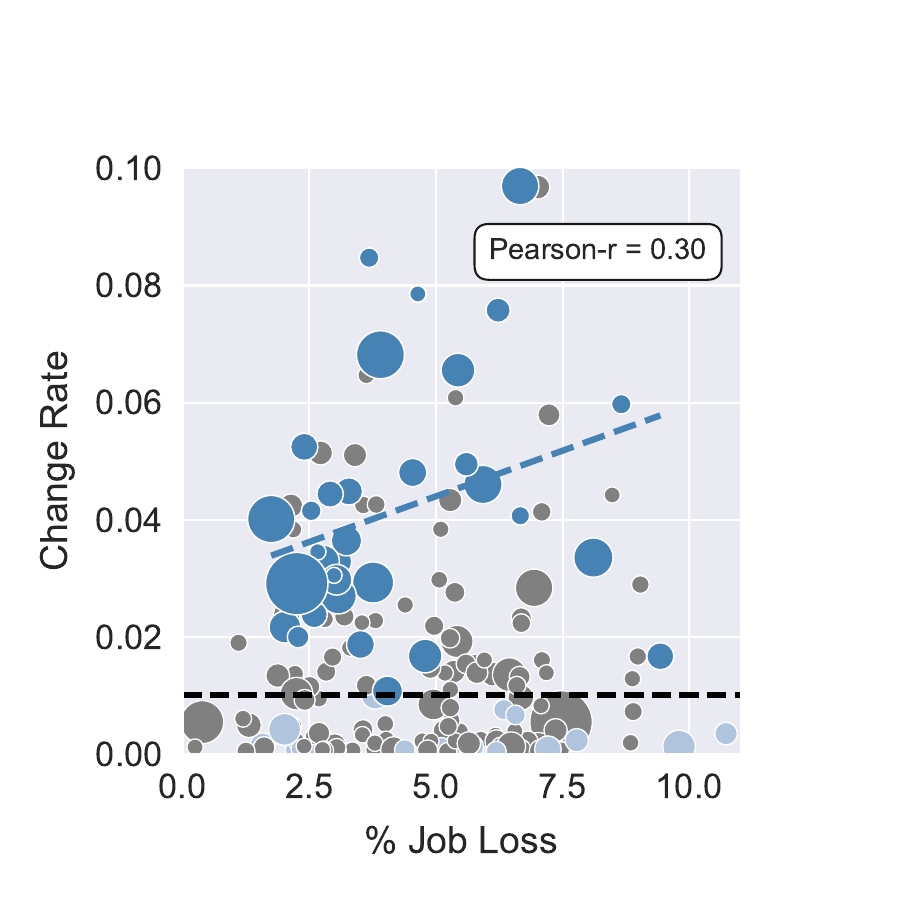}}
    \subfloat[Asian]{\includegraphics[width=0.25\linewidth]{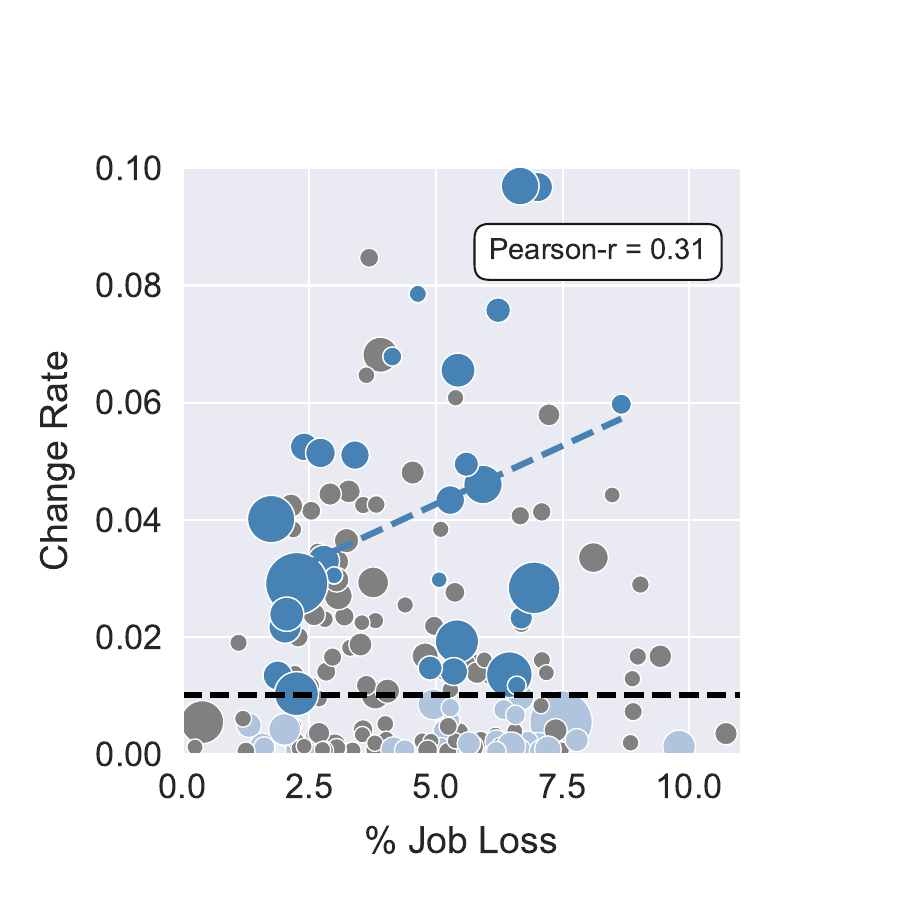}}
    \subfloat[Hispanic]{\includegraphics[width=0.25\linewidth]{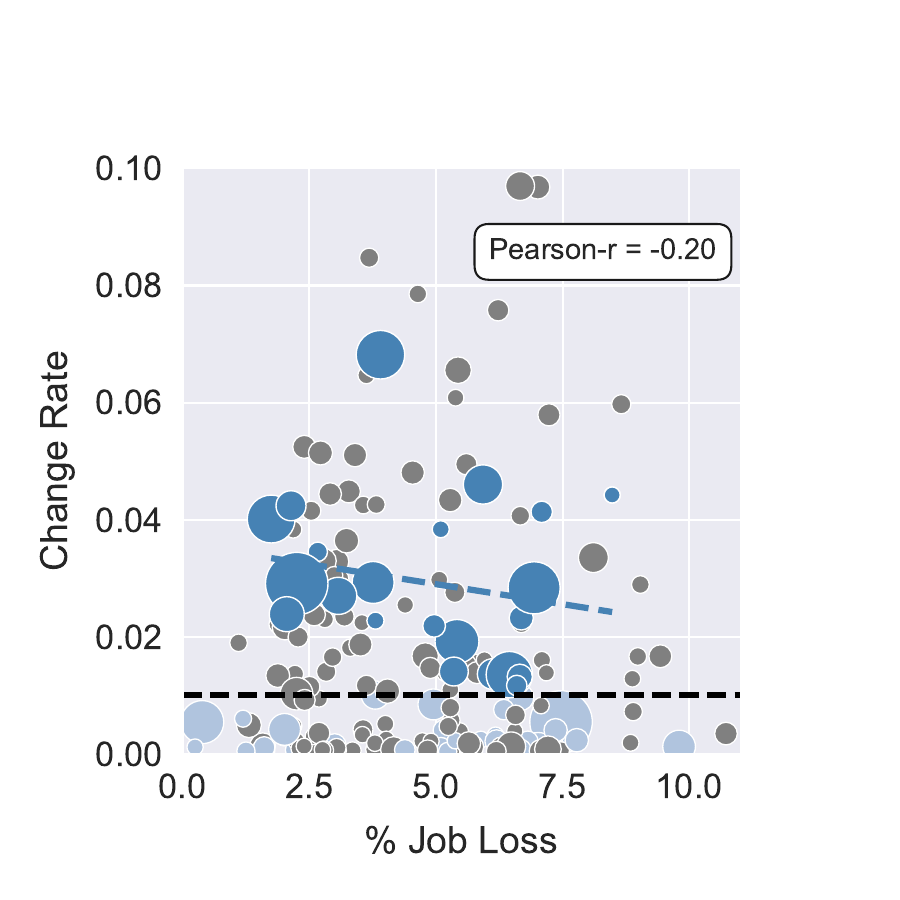}}
    \caption{County-level results for job loss due to the pandemic by August 2021.}
    \label{fig:county_job}
\end{figure}

\begin{figure}[htbp]
    \centering
    \subfloat[Black]{\includegraphics[width=0.25\linewidth]{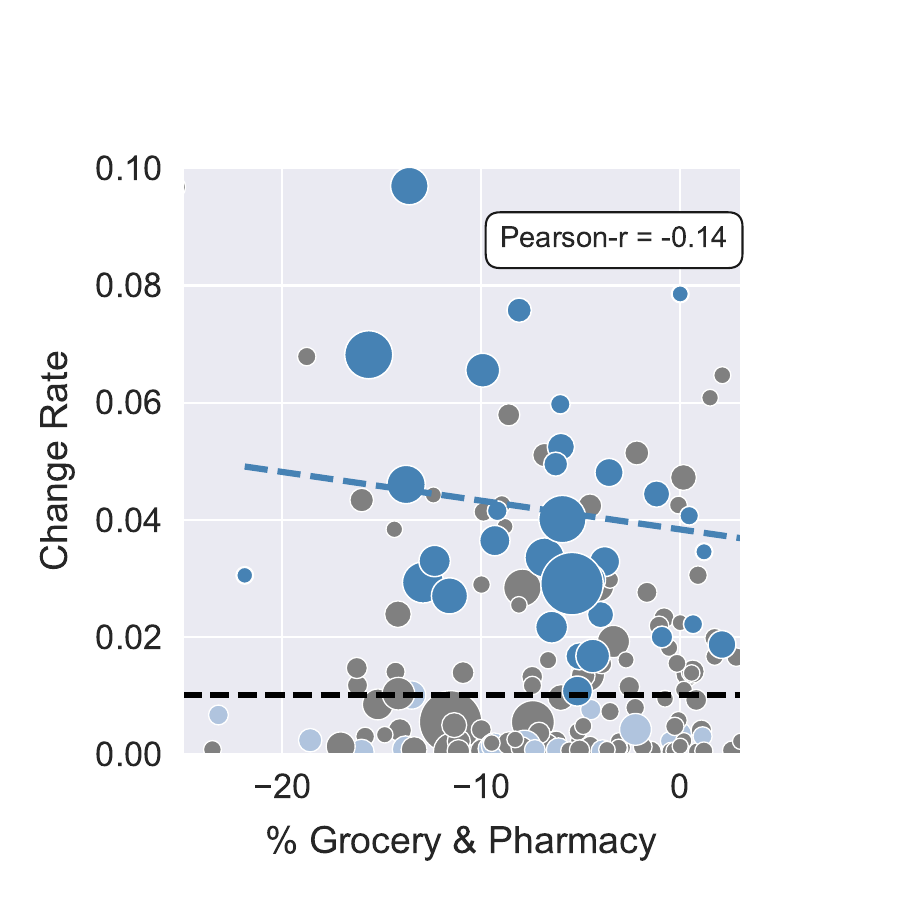}}
    \subfloat[Asian]{\includegraphics[width=0.25\linewidth]{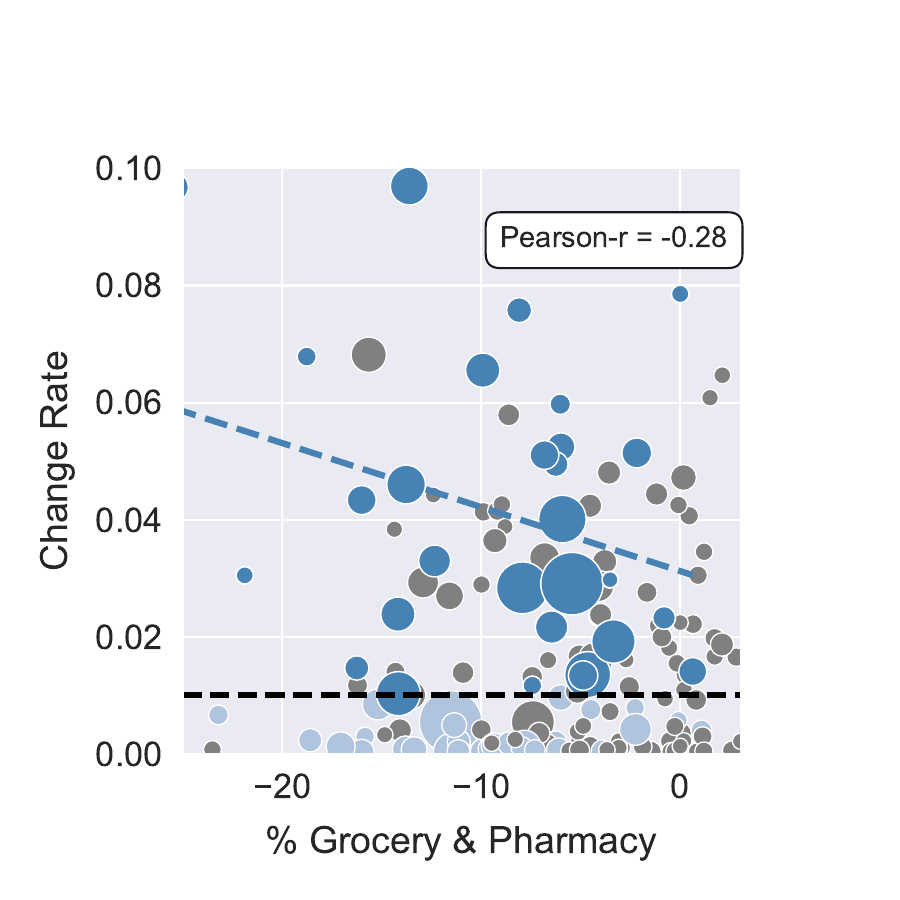}}
    \subfloat[Hispanic]{\includegraphics[width=0.25\linewidth]{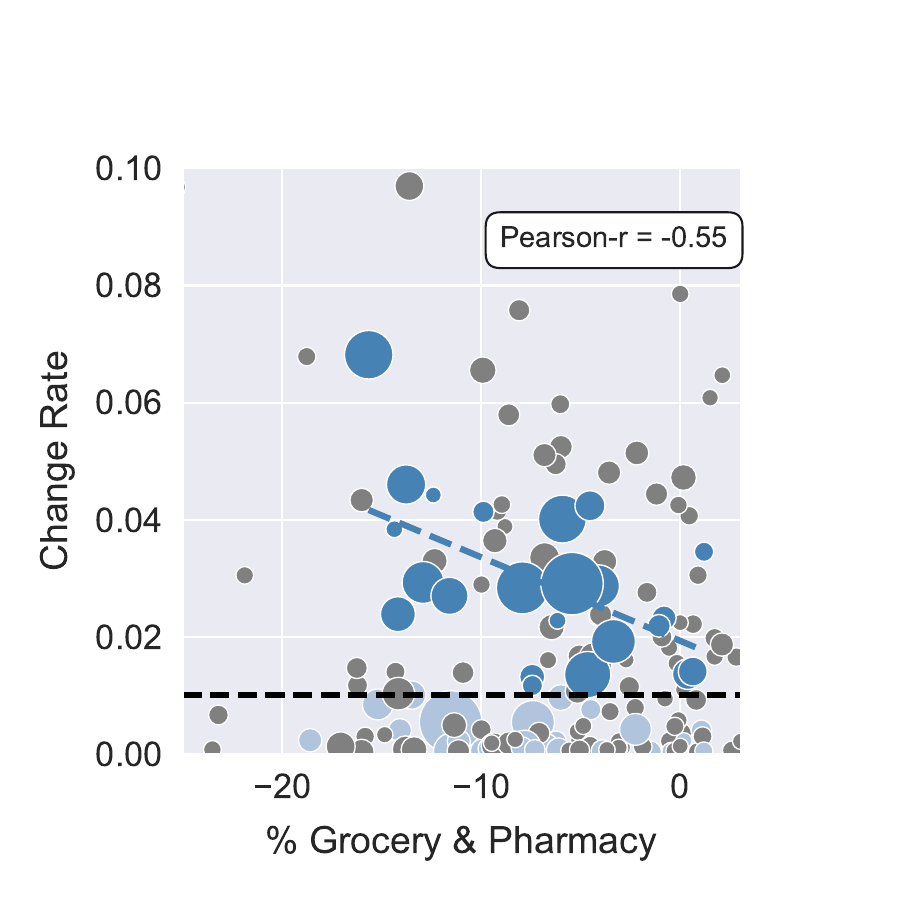}}\\
    \subfloat[Black]{\includegraphics[width=0.25\linewidth]{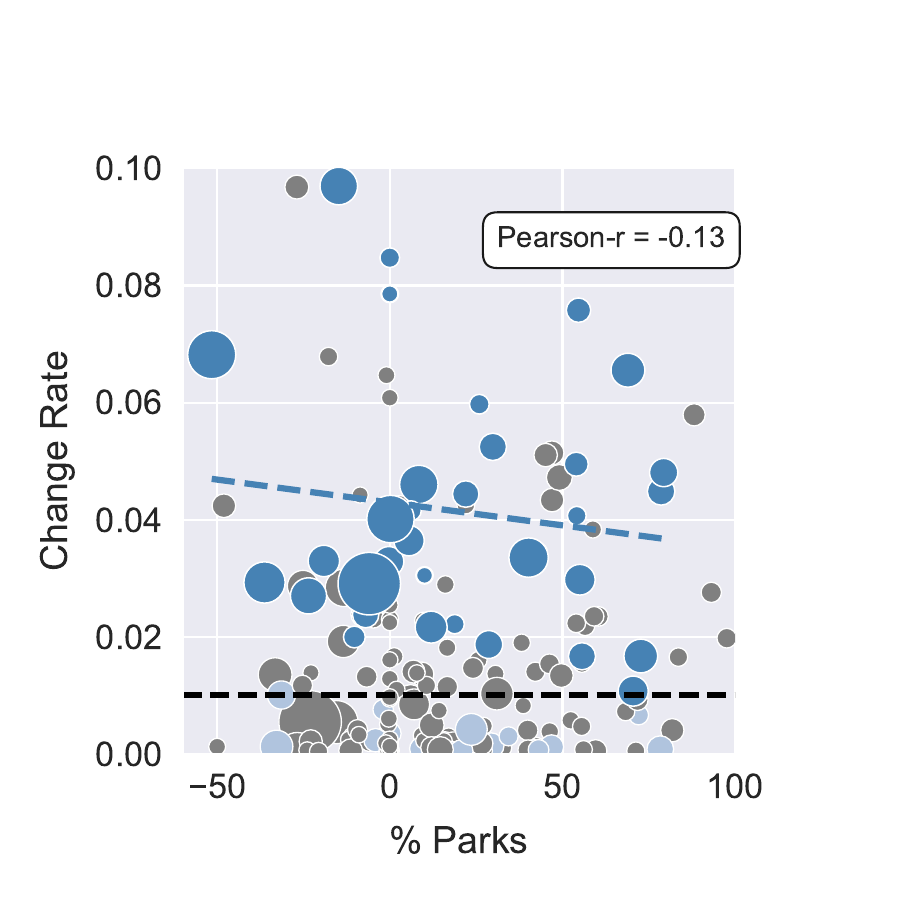}}
    \subfloat[Asian]{\includegraphics[width=0.25\linewidth]{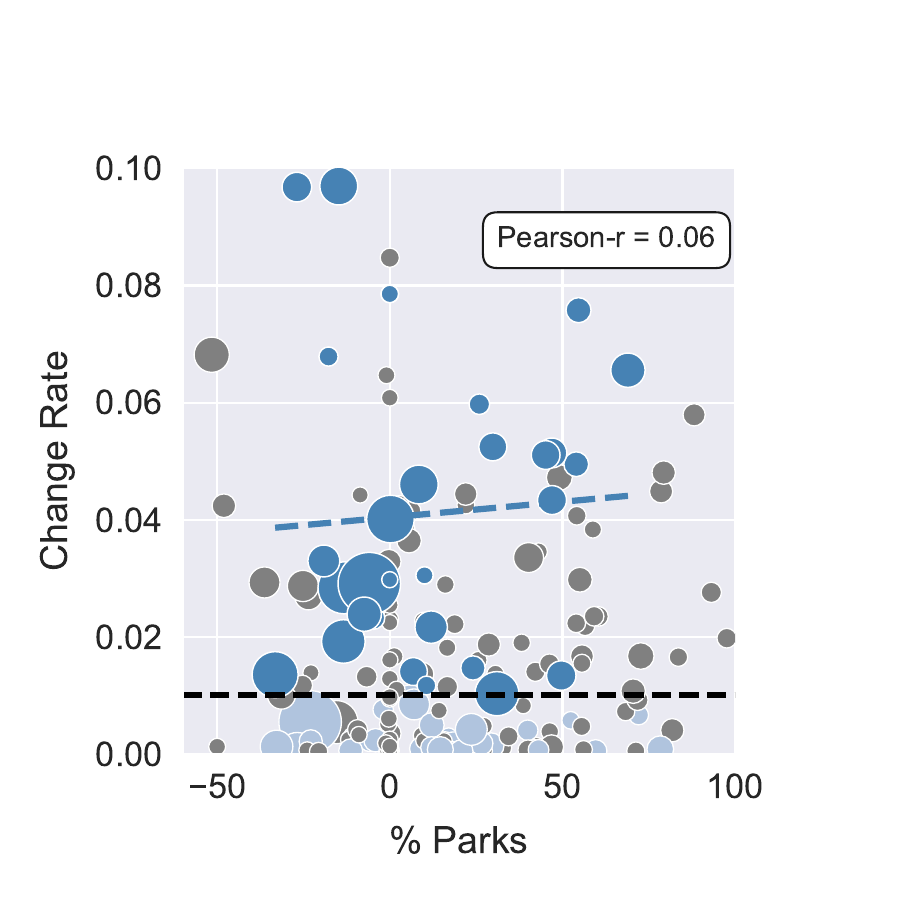}}
    \subfloat[Hispanic]{\includegraphics[width=0.25\linewidth]{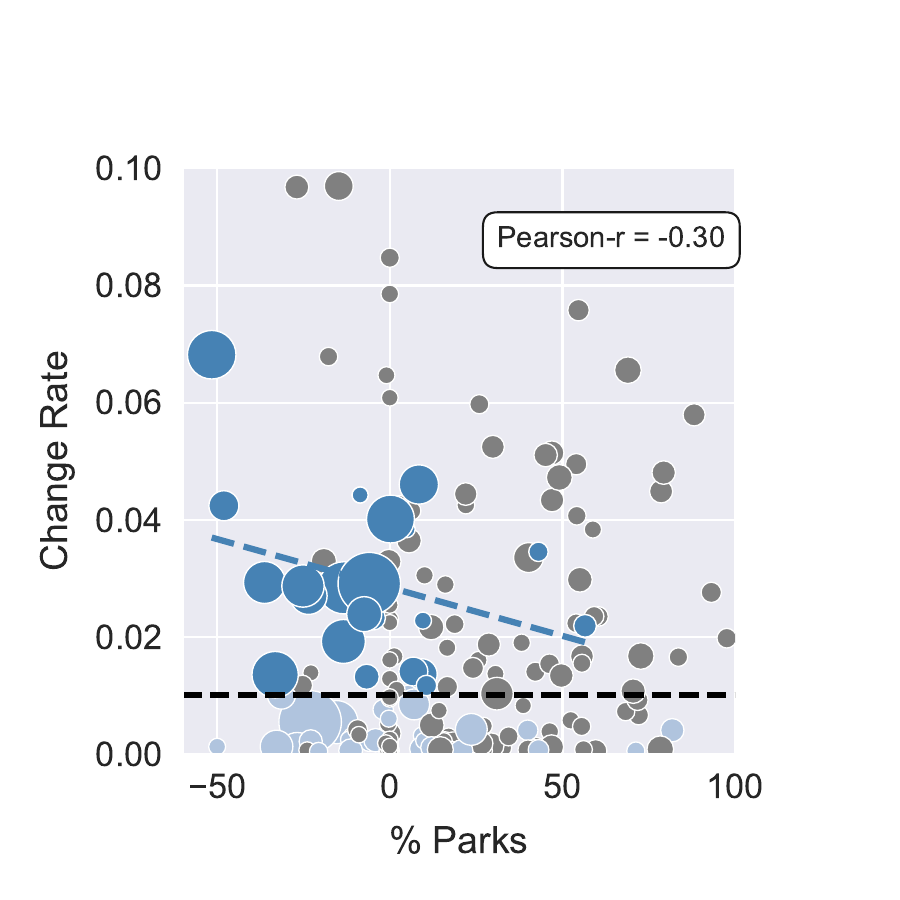}}
    \caption{County-level results for mean shifts in mobility activities by the end of June 2020.}
    \label{fig:county_mobility}
\end{figure}

We further check if the observed racial and ethnic disparities of transit supply reduction may also correlate to pandemic-related health, economic and mobility consequences for the impacted transit-dependent communities. This helps us to explore the potential concurrent implications of such disparities in relation to the well-being and essential mobility intensity of the impacted communities. To this end, we focus on the relationships between the change rate and the health, economic and mobility factors in the most impacted areas with a high density of certain racial-ethnic groups. For health-related indicators, we use the data on COVID-19 deaths, available on the county level for all 381 covered counties by 232 transit agencies~\citep{nyt_covid}. Economic impacts are based on the expected job loss rate as a result of the pandemic~\citep{pandemic_job_loss_data} and mobility changes are captured by the shifts in activities related to grocery \& pharmacy, and park places~\citep{google_mobility}. Here, we focus on the counties rather than CTs due to the aforementioned data being only available at the county level, and we also aggregate the change rate of transit services to the associated counties. As for the most impacted racial-ethnic communities, we target the communities corresponding to the top 25\% of the minority compositions in the transit-covered communities that suffer the top 25\% change rate in transit supply. The resulting number of such counties for Black, Asian and Hispanic groups are 34, 29 and 23, respectively. The analyses for the White group are excluded in this section due to only one county (Pitkin, CO) meeting the above criteria. \Cref{fig:county_covid} to \Cref{fig:county_mobility} show the results of the relationships between the change rate and cumulative COVID-19 death rates, job losses and mobility intensities, where the most-impacted counties are highlighted in blue and the dot size represents each county's number of population. The dashed horizontal line shows the threshold for the 25th percentile of the change rate in all counties. The Pearson correlation coefficients (r) between the change rate and the selected factors for the most impacted counties are also displayed. As can be observed in \Cref{fig:county_covid}, there is a strong correlation, ranging from 0.40 to 0.52, between the loss in transit service and the COVID-19 death rate for all three racial-ethnic groups and the correlations are all statistically significant (\(p < 0.10\)). In other words, impacted counties with a high density of minorities that experience a higher rate of transit loss also happen to suffer more from COVID-19 deaths. While such correlations are not significant for other less impacted communities, the results highlight the ripple effects of transit shortage for the most vulnerable groups.  Moreover, the relationship between the job loss rate and the change rate points out a difference among the three racial-ethnic groups (see \Cref{fig:county_job}). The only group with a positive and significant correlation is the Black group (\(r(32) = 0.30, p < 0.10\)). On the other hand, the association between the change rate and job loss due to the pandemic is positive but insignificant for Asian Americans (\(r(27) = 0.31, p > 0.10\)), and negative but insignificant for the Hispanic group (\(r(21) = -0.20, p > 0.10\)). Finally, the results for the shifts in essential mobility activities related to grocery \& pharmacy, and park categories are shown in \Cref{fig:county_mobility}. Both trends suggest a more prominent negative correlation between the change rate and visits to essential activities for Hispanics as compared to other groups. The correlation between the grocery and pharmacy mobility with change rate for this group is strongly negative and statistically significant (\(r(21) = -0.55, p < 0.10\)) while the correlation between the parks mobility activities and change rate is negative but statistically insignificant (\(r(21) = -0.30, p > 0.10\)). All the other correlations for Black and Asian groups are found to be statistically insignificant (\(p > 0.10\)), suggesting less substantial shifts in the essential mobility activities for the highly impacted counties with a high composition of Black and Asian Americans by the end of June 2020.

To summarize, analyzing the correlation between pandemic-related burdens and the service change rate in transit-dependent counties suggests that the loss of transit supply may have ripple effects on socioeconomic and mobility activities that vary among different racial/ethnic groups, especially in the most impacted communities. These observations verify our final hypothesis \textbf{H-4}. We note that death rates and mobility intensities are related to the more immediate aftereffects of the COVID-19 outbreak as they capture the cumulative number of death cases and mean changes in mobility activities by the end of June 2020. On the other hand, the job loss rate pertains to the subsequent impacts of the pandemic, signifying the percentage of jobs lost by August 2021. As such, we believe that immediate mobility changes for the Hispanic group exhibit a stronger correlation with the change rate, whereas the enduring economic challenges of the pandemic correlate most with the Black group. Additionally, all three economic and mobility indicators related to the Asian group were found to be insignificant, which may indicate better access to other mobility solutions for Asians as compared to the two other groups. Finally, note that the above conclusions are in line with our results in the previous sections, as a disparity in transit loss was revealed, with Black Americans being the most impacted group while less impact on the Asian and White groups was identified. 


\section{Summary and Conclusion}
\label{section:discussion}
\renewcommand{\arraystretch}{2}
\begin{table}[H]
\caption{Summary Results}
\centering
\footnotesize
\label{table:summary_results}
\begin{tabularx}{\textwidth}{@{} p{0.33\textwidth} *{4}{X} @{}}
\toprule
& \textbf{White} & \textbf{Black} & \textbf{Asian} & \textbf{Hispanic} \\
\midrule

Impact on pre-pandemic supply & $-0.083^{***}$  & $0.094^{***}$ & $-0.098^{***}$ & $-0.034^{***}$ \\[2.5pt] 

Direct effects on change rate & $0.002^{*}$ & $-0.005^{***}$ & $-0.001$ & $-0.002^{*}$ \\[2.5pt]

Indirect effects on change rate & $-0.020^{***}$ & $0.019^{***}$ & $-0.025^{***}$ & $-0.007^{***}$ \\[2.5pt]

Agency-level indirect effects on change rate & negative for $90$\% of agencies ($>20$\% significant$^{.}$) & positive for $88$\% of agencies ($>40$\% significant$^{.}$) & negative for all agencies ($>50$\% significant$^{.}$) & negative for $90\%$ of agencies agencies ($>50$\% significant$^{.}$) \\[2.5pt]

Agency-level direct effects on change rate & positive for all agencies ($>50$\% significant$^{.}$) & negative for all agencies ($>95$\% significant$^{.}$) & negative for $93$\% of agencies (none significant) & negative for all agencies ($>70$\% significant$^{.}$) \\[3.5pt]

Correlation with pandemic death rates (most impacted)& $-$ & $0.43^{.}$ & $0.52^{.}$ & $0.40^{.}$ \\[2.5pt]
Correlation with pandemic-induced job losses (most impacted) & $-$  & $0.30^{.}$ & $0.31$ & $-0.20$ \\[2.5pt]
Correlation with shifts in grocery\& pharmacy activities (most impacted)& $-$ & $-0.14$ & $-0.28$ & $-0.55^{.}$ \\[2.5pt]
Correlation with shifts in parks activities (most impacted)& $-$ & $-0.13$ & $0.06$ & $-0.30$ \\[2.5pt]
\bottomrule
\multicolumn{5}{l}{\scriptsize{$^{***}p<0.001; ^{**}p<0.01; ^{*}p<0.05; \Dot{}p<0.1$}}
\end{tabularx}
\end{table}

In this study, we investigate the racial-ethnic disparities in the reduction of transit supplies as a result of the COVID-19 pandemic. Through the integration of causal mediation analysis applied to data collected from 232 agencies spanning 45 states, we establish connections between the racial-ethnic compositions of communities, pre-pandemic supply levels, and shifts in transit service, all while accommodating the heterogeneous capabilities of the selected agencies. Our findings show that transit-dependent and lower-income communities and certain racial-ethnic groups have experienced disproportional reductions in their absolute reduction of transit supply. Furthermore, accounting for the pre-pandemic level of supply reveals the disparities in the reduction of transit service to more affect Black and Hispanic communities, compared to White and Asian groups. \Cref{table:summary_results} presents a summary of our main findings across the four target racial-ethnic groups. In short, our comprehensive analyses led to four major conclusions given the four target racial-ethnic groups, as summarized below:

\begin{itemize}
     \item A higher composition of White Americans in CTs was associated with less reduction in the quantity of transit service during the pandemic. Nevertheless, the percentage of the White population was also found to have a negative relationship with the pre-pandemic transit supply (\(p<0.001\)). When controlling for the initial supply level, significant and negative total effects on the transit service reduction for the White group were observed at the 0.001 level (Total Effects = $-0.018$ vs. Black Total Effects = $0.014$). These findings favor the conclusion that the White group, along with the Asian group, are less impacted by the disruption in transit service as compared to other racial-ethnic groups.

    \item Our analysis indicated that among the target groups, Black Americans were the most reliant on transit. They were the only racial-ethnic group with a positive relationship with pre-pandemic transit supply levels (\(p<0.001\)). But they also suffered the highest loss in absolute service reduction during the pandemic. Our analysis of their transit losses suggests that their high pre-pandemic transit dependency (as a mediator) resulted in the highest negative total effect on the reduction of transit services (ACME = $0.019$, $p<0.001$). This direct loss was also shown to be strongly (\(p<0.1\)) associated with pandemic-induced job loss and COVID-19 death rate among the Black population. Nevertheless, we also observed a negative direct effect (ADE=$-0.005$, \(p<0.001\)) for the Black population, which implies efforts from the majority of the transit agencies to maintain their level of transit service that may have prevented an additional 27\% loss in the transit supply.

    \item For the Asian group, similar to White Americans, we report a negative relationship with transit supply prior to the pandemic (\(p<0.001\)). This group also displayed a most negative indirect effect (ACME=$-0.025$, \(p<0.001\)) on transit supply reduction between the target groups, while their direct effect on the change rate was insignificant. Additional analysis indicated potential indifference toward their transit needs during the pandemic, as no significant direct effect was observed across any agencies at the 0.1 level. Furthermore, correlations between the transit service loss and indicators related to job losses and essential mobility shifts were insignificant for this group, suggesting that Asian Americans were less affected by transit loss due to the pandemic compared to other minority groups. 

    \item Our findings suggest that the Hispanic group is the second most affected racial-ethnic group after the Black group. We observed Hispanic Americans to be less transit-dependent and disproportionately affected by the quantity of service loss compared to the Black group. As a result, despite having a negative and significant direct effect (ADE=$-0.002$, \(p<0.05\)) on the transit supply reduction (similar to the Black group), their total effect was notably lower and significant (Total Effect = $-0.009$ vs. Black Total Effects = $-0.024$). Regarding the difference between the Hispanic and Black groups, further analysis revealed that the Hispanic group exhibited a stronger correlation with shifts in immediate pandemic consequences, such as visits to grocery and pharmacy locations (Pearson-r = $-0.55$). On the other hand, the Black group demonstrated a stronger association with more lasting pandemic implications like higher rates of job loss (Pearson-r = $0.30$).
    \end{itemize}

As public transportation remains an essential mobility service for vulnerable populations and disadvantaged communities in the post-pandemic era, transit agencies face the challenge of adapting to the ``new normal'' where substantial modifications in their practices are required~\citep{berrebi2021bus, speroni2023pandemic}. Under the raised complexities, the knowledge and insights generated in this study will have significant implications for the restoration of transit services and highlight the pathways for local and federal transit agencies to retain equitable transit services for the general public. Nevertheless, there are several additional aspects that warrant consideration. First, even prior to the pandemic, transit ridership declines were widespread in U.S. cities~\citep{erhardt2022has}. For instance, the year 2019 saw U.S. bus ridership at its third-lowest since World War II, following a trend of seven consecutive years of reduction ~\citep{NAP25635, berrebi2021bus}. At the same time, a review of transit agencies' operations suggested a prevailing focus on serving more affluent riders and voters as a whole rather than prioritizing vulnerable populations~\citep{taylor2015public}. Therefore, we point out that the disparities in transit service, where specific racial-ethnic groups experience disproportionate impacts from supply changes, likely extend beyond the scope of the pandemic era. Second, our approach not only underlines the complexities involved in identifying the disparities within communities as a result of transportation policies but also provides a broad overview of the racial-ethnic disparities during the pandemic. However, given the multifaceted nature of this challenge, additional targeted inquiries are imperative. For instance, linking the shared patterns among agencies to their operational and strategic capacities or establishing causal relationships between pandemic-related service adjustments and economic and health consequences demands further exploration. In this context, case studies focusing on specific regions and agencies can provide a microscopic perspective that complements the broader analysis. Lastly, our study makes a substantial contribution to the growing body of evidence highlighting the disproportionate impact on vulnerable populations during times of crisis~\citep{neal2020economic, wheelock2020comparing, reed2012race}. This observation holds significant implications, as our perspective can be extended to both past and potential future nationwide emergencies, shedding light on the consequences of our policy responses. Through this lens, we gain valuable insights into strategies for restoring the resilience of public infrastructure in the long term to better protect vulnerable communities.

\printcredits

\bibliographystyle{ieeetr}

\bibliography{references}


\end{document}